\documentclass[twocolumn,showpacs,showkeys,preprintnumbers]{revtex4}
\usepackage{graphicx}
\usepackage{subfigure}
\usepackage{lscape}
\usepackage{dcolumn}
\usepackage{bm}
\usepackage{color}

\begin{document}

\title{Searching for correlations of geomagnetic activities with high energy EAS muons}
\author{Rajat K. Dey}
\email{rkdey2007phy@rediffmail.com}
\author{Sabyasachi Ray}
\email{methesabyasachi@gmail.com}
\author{Sandip Dam}
\email{sandip_dam@rediffmail.com}
\affiliation{Department of Physics, University of North Bengal, Siliguri, WB 734 013 India}

\begin{abstract}
The paper aims to explore the asymmetry of the muon content of non-vertical and very high energy Monte Carlo showers due to the influence of Earth's geomagnetic field. Simulations have shown that the geomagnetic field modifies the trajectories of muons in a shower producing a polar asymmetry in the density/number of positive and negative muons in the shower front plane. The asymmetry is quantified by a transverse separation between the positive and negative muons barycentric positions through opposite quadrants across the shower core. The dependence of this transverse muon barycenter separation (TMBS) on polar position shows a clear maximum at a position that is correlated with the primary composition and geomagnetic activities. It is noticed that the maximum TMBS parameter exhibits sensitivity to any transient weakening of Earth's magnetic shield caused by geomagnetic storm originated from bursting solar processes. Obtained simulation results are quite important to design any possible new experiment based on these features of muons in extensive air showers. 
\end{abstract}

\pacs{96.50.sd, 95.75.z, 96.50.S} 
\keywords{cosmic-ray, EAS, geomagnetic activity, muons, simulations}
\maketitle

\section{Introduction}

Analyzing directly accessible data from EAS experiments one or more parameters sensitive to the mass composition of cosmic rays (CR) are usually measured [1]. Measurements of these parameters are made either by individual or hybrid detection techniques. The observation of an EAS provides electron and muon lateral density distribution (LDD) data at the observational level along with their timing information. The simulated/observed LDD data are usually reconstructed by means of a lateral density function \emph{e.g.} Nishimura-Kamata-Greisen (NKG) structure function [2] in order to get the global EAS parameters. The lateral distribution of cascade particles in an average EAS is often assumed to be symmetrical in the plane perpendicular to the shower axis. However, due to the intrinsic shower-to-shower fluctuations and geomagnetic effects, the assumed axial symmetry would have been perturbed noticeably for highly inclined showers ($\Theta \geq 50^{\rm o}$). Such effects may even distort the axial symmetry to the distribution of EAS muons even in vertically incident showers. Inclined showers though experience similar effects as vertical showers but can exhibit significant asymmetries.

The shower data analysis is usually performed by projecting the ground detector signals onto the shower front plane without accounting the further shower evolution of the late regions. As a consequence the circular symmetry in the LDD data obtained from ground detectors is broken for inclined showers. The LDD data exhibit polar asymmetry in the shower plane due to the different amount of atmosphere traversed by the shower particles. These are well-known geometrical and attenuation effects to polar asymmetries of an EAS [3]. To establish a correlation of Earth's geomagnetic activity with the magnitude of asymmetry of the LDD of muons only, the geometric and attenuation effects must be eliminated in shower data analysis. In the first instance, our data analysis technique will remove/correct the polar asymmetry caused by the geometric effect. Next, the asymmetry arises from the attenuation effect would be ignored judiciously as muons suffer small attenuation in the atmosphere. The present article aims to get some fingerprints of the geomagnetic field (GMF) in polar distribution of high energy muons, and to quantify the asymmetric distribution caused by the GMF in terms of a parameter, called the transverse muon barycenter separation (TMBS). The TMBS parameter is defined as the linear distance between the barycenter positions of positive and negative muons in the shower front plane. 
\begin{figure*}[ht]
\subfigure
{\includegraphics[scale=0.35]{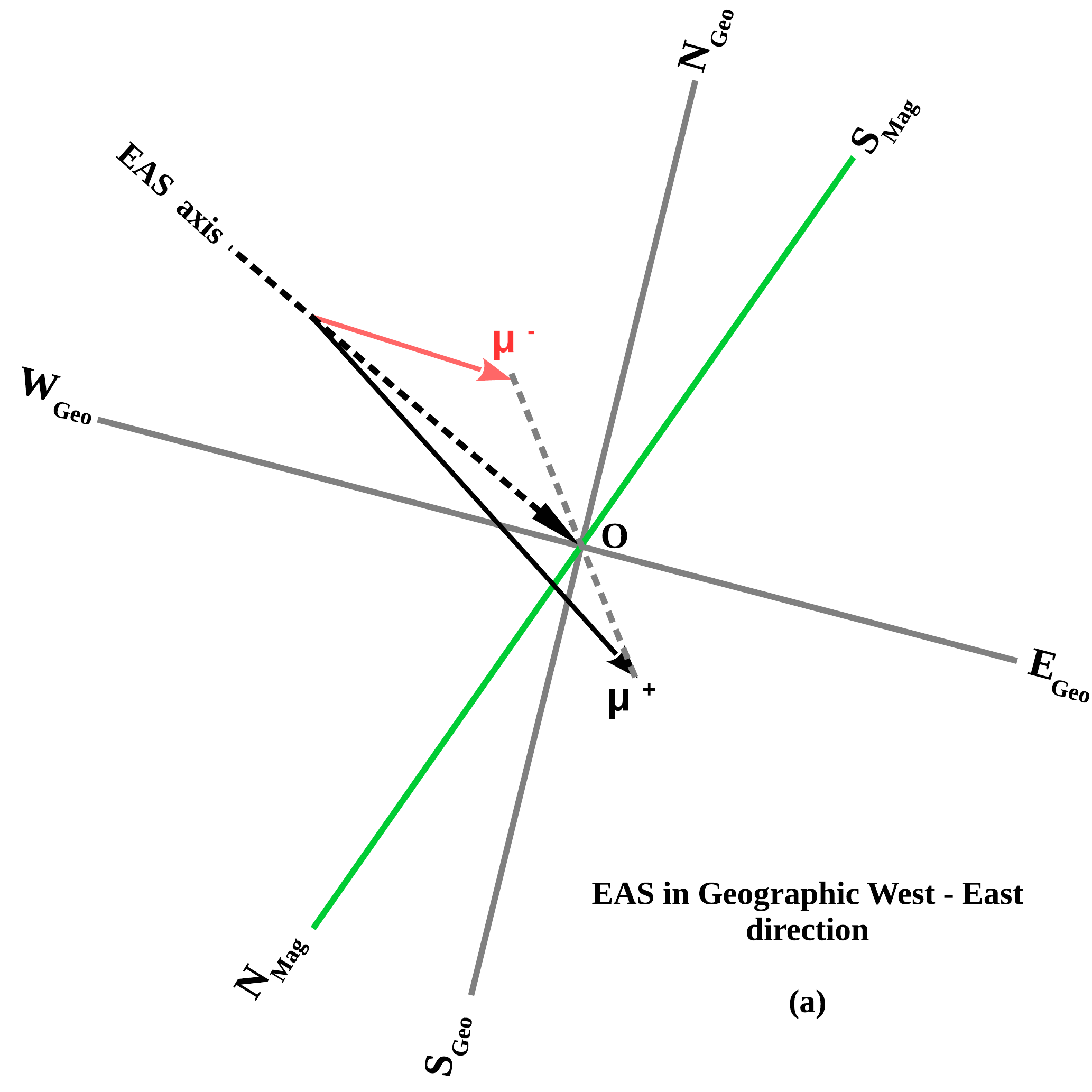}}
\subfigure 
{\includegraphics[scale=0.35]{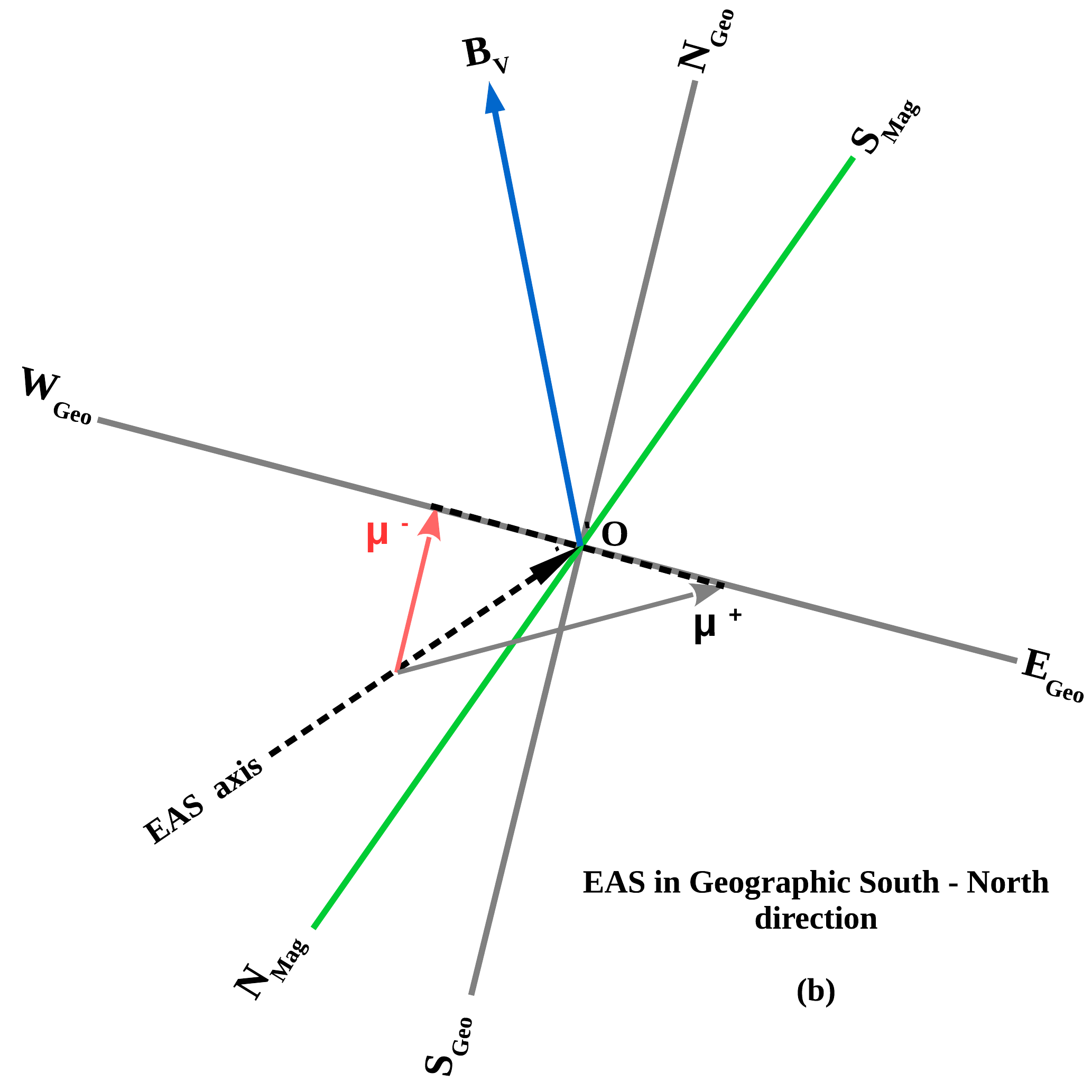}}
\caption{The separation of $\mu^{+}$ and $\mu^{-}$ generated from a parent particle in an EAS experiencing the GMF at a place with east magnetic declination. The pair of lines represent the geographic and magnetic meridians respectively.}
\end{figure*} 
When a shower arrives at an arbitrary angle to the GMF, the charged particles in the EAS experience deflections. In the cascade, the soft component i.e. electrons have short radiation lengths interacting strongly in the atmosphere. They also suffer stronger bremsstrahlung effects, thereby rapidly changing the directions of their momenta relative to the GMF. These processes cause wider lateral spread of electrons in the EAS, while the effect of GMF on them is insignificant. Muons travel large distances through the atmosphere encountering negligible scattering and hence come under the influence of GMF significantly [4]. Furthermore, some recent studies indicated that the geomagnetic effect became more pronounced on medium to high momenta muons than lower ones, particularly for showers with primary energy, $\rm{E} \geq 1$ PeV and zenith angle, $\Theta \geq 50^{\rm o}$ [5-7]. Fig. 1 demonstrates the deflections of a positive muon ($\mu^{+}$) and a negative muon ($\mu^{-}$) by a GMF generated from a parent particle in a shower arriving from two different directions. The Fig. 1a shows the deflections of $\mu^{+}$ and $\mu^{-}$ in opposite directions generated from a CR particle coming from the geographic west and advances into the east direction. But, the Fig. 1b shows a situation where the EAS comes from the geographic south and advances into the north direction. Some earlier Monte Carlo (MC) simulation studies reported that the CR mass composition can be determined from the fingerprints of lateral muon distribution and the muon charge ratio (the ratio of $\mu^{+}$ to $\mu^{-}$ numbers) at convenient distances from the shower core [5-11].

In this paper the magnitude of the asymmetry arising from the influence of the GMF on EAS muons is investigated via air shower simulation, using the code {\emph{CORSIKA}} [12]. To quantify the asymmetry it is necessary to estimate the expected TMBS parameter for each shower at different polar positions in the shower plane. Next, the maximum value of the TMBS parameter (called MTMBS) is estimated from the variation of TMBS parameter with polar angle in order to correlate the geomagnetic activity with high energy muons in an EAS. The work also discusses the impact of Earth's magnetic field components on the MTMBS parameter at certain geographical places. Finally, sensitivities of the MTMBS parameter to CR mass composition and to any transient weakening of Earth's magnetic shield are demonstrated.

In the following, we describe the {\emph{CORSIKA}} simulations in section 2. Section 3 describes the data analysis procedure for the estimation of the TMBS and MTMBS parameters. The results are presented and discussed in section 4. Section 5 provides a possible experimental approach and some potential ideas for further studies. This section also summarizes our conclusions.

\section{Monte Carlo simulations}
The \emph{CORSIKA} code [12] was used to simulate the EAS evolution in the atmosphere. The MC showers are generated using UrQMD [13] at low energies ($\rm{E_{h}} < 80 {\rm{GeV/n}}$) combined with EPOS $1.99$ [14] as the model in the high energy regime for the mechanisms of hadronic interactions. We consider the U.S. standard atmospheric model as parametrized by J. Linsley in the simulation [12]. The physics of the electromagnetic (EM) interactions was dealt with the EGS$4$ program library [15]. 

The MC showers have been generated at the altitude and geomagnetic fields corresponding to the KASCADE air shower array [16]. It is of interest to observe the nature of variation of the MTMBS	parameter with geomagnetic field components at some specific locations. Hence, two sets of specific locations have been selected in the simulations to generate more showers. Locations in set I (Talon, Khabaresvsk Republic, Sakha Republic, Aldanskya Sakha Republic and Nyuya) have the same latitude but a variable longitude. Set II (Angola, Republic of Congo, Cameroon, Chad and Libya) follows exactly opposite settings i.e. nearly the same longitude but a variable latitude. In Table I and II, we have given a detailed latitude and longitude data of these locations in set I and II respectively. Using information available from NOAA [17], altitude and GMF components have been set accordingly in CORSIKA steering
files [12].  

The MC showers have been generated at a fixed 1 PeV primary energy, and for some specific studies with a spectrum following a power law with $\rm{E}^{-2.7}$. A considerable amount of MC showers are simulated in three limited primary energy regions: $1-3$, $8-12$ and $98-102$ PeV. On observation levels, the kinetic energy cut-offs are set as $0.003$ GeV for electrons and $0.3$ GeV for muons. MC showers are generated for proton (p) and iron (Fe) primaries with roughly the same statistics. The percentage of ${\mu}^{\pm}$ generally increases with $\Theta$ over $\rm{e}^{\pm}$ in a shower. Shower simulations are therefore done with $\Theta \geq 50^{\rm o}$ so that the GMF be more effective on muons. No restriction was applied for the azimuthal angle ($\Phi$) in simulations. To examine the effect of the GMF distinctly, showers are also simulated by switching off Earth's magnetic field. 

\begin{table*}
\begin{center}
\begin{tabular}
{|l|l|l|l|l|r|} \hline
{\rm{Location}} & {\rm{Latitude (deg)}} & \rm{Longitude (deg)} & $\rm {B_{H}}({\mu}\rm{T})$ & $\rm{B_{V}}(\mu \rm{T})$ & $\rm{B}(\mu \rm{T})$ \\ \hline

Talon & 60.41385  & 148.02429 & 16.9138 & 54.1537  & 56.7336 \\ 
Khabaresvsk rep. & 60.75916  & 139.76257 & 15.917  & 55.977  & 58.0214\\
Sakha rep. & 60.23981  & 130.44616 & 15.1874 & 57.3055 & 59.2839 \\
Aldanskya Sakha rep. & 60.15244 & 124.29382 & 14.4524  & 58.3457  & 60.109  \\
Nyuya  & 60.06484 & 116.20789  & 13.506 & 59.4667  & 60.981 \\ 
\hline   							
\end{tabular}
\caption {Details of the GMF components of some geographical locations with nearly constant latitude.} 
\end{center}
\end{table*} 

\begin{table*}
\begin{center}
\begin{tabular}
{|l|l|l|l|l|r|} \hline
{\rm{Location}} & {\rm{Latitude (deg)}} & \rm{Longitude (deg)} & $\rm {B_{H}}({\mu}\rm{T})$ & $\rm{B_{V}}(\mu \rm{T})$ & $\rm{B}(\mu \rm{T})$ \\ \hline

Angola & $-11.5230$  & 16.907949  & 19.6794 & -124.6477 & 31.5402\\ 
Rep. of Congo & $-1.93323$  & 16.715698  & 27.7198  & -17.729  & 32.9045 \\
Cameroon & $7.1881$   & 16.715698  & 33.4432  & -5.8773  & 33.9557  \\
Chad & $15.79225$   & 16.715698  & 35.2796  & 7.6818  & 36.1238  \\
Libya  & $21.12549$   & 16.715698  & 34.6308 & 16.0294 & 38.1606\\ 
\hline   							
\end{tabular}
\caption {Details of the GMF components of some geographical locations with nearly constant longitude.} 
\end{center}
\end{table*} 
      
\section{Description of the data and selection cuts}

To retain only the GMF effect in the polar distribution of high energy muons, the following steps have to be implemented in shower data analysis.

For inclined showers, the gross muon data ($\rho_{\mu}$ or $N_{{\mu}^{\pm}}$) get overestimated in the early region while underestimated in the late region of an EAS. Alternatively, it means that muons in the late region do undergo higher attenuation than those arriving in the early region in the shower. This feature in the evolution of an EAS accounts the attenuation contribution to the overall polar asymmetry of the distribution of muons. This is the well-known attenuation effect in the cascade theory. The polar asymmetries are corrected out from gross EAS data when the information of muons is being transformed from the ground plane to the more justified shower front plane, and is termed as the geometric effect.

\begin{figure*}[htbp]
\centering
\includegraphics[width=0.85\textwidth,angle=0]{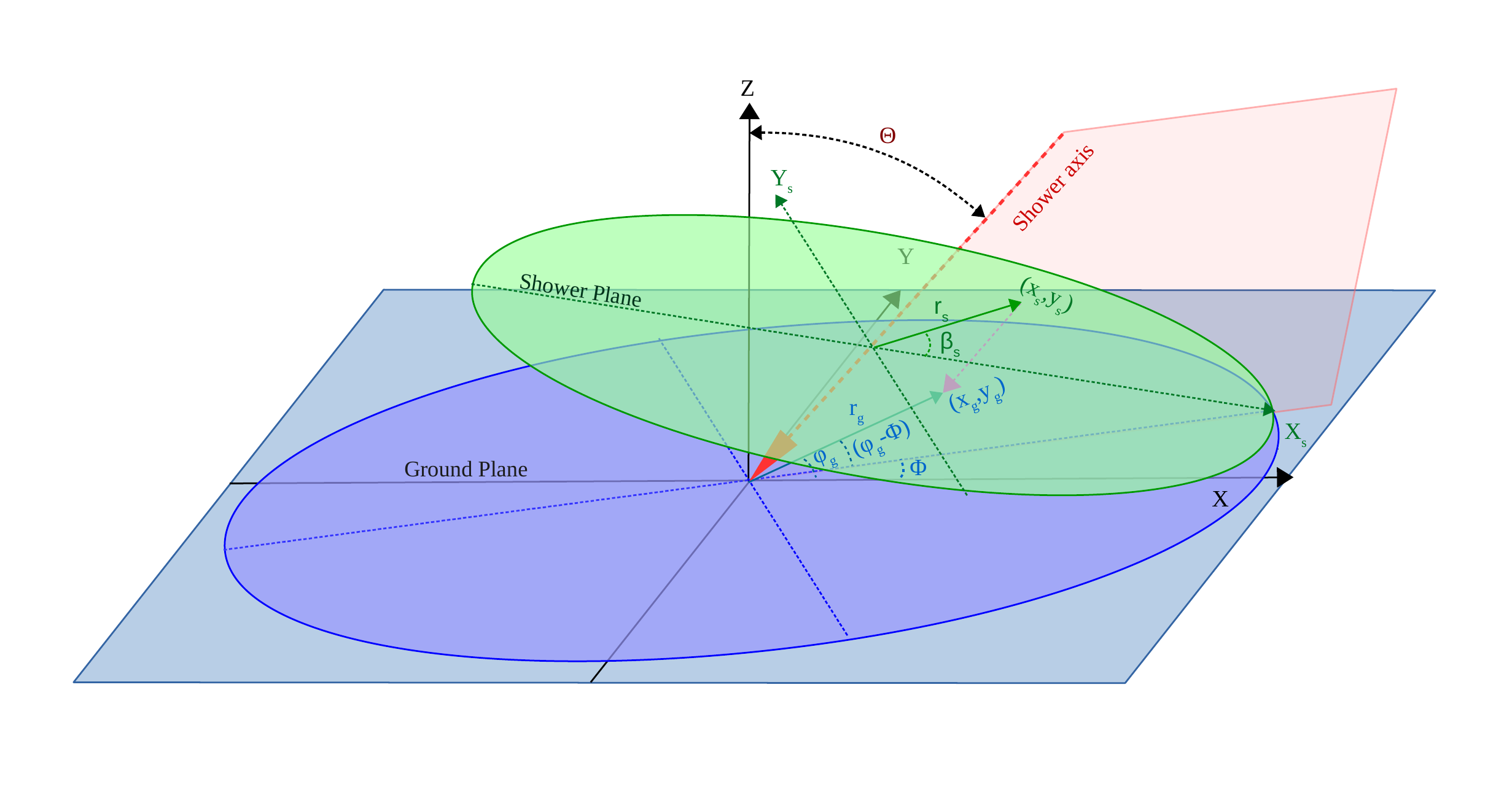}\vfill
\caption{Geometry of the ground plane and shower front plane used for the geometric correction in an inclined shower. $\Theta$ and $\Phi$ represent the shower zenith and azimuth angles.}
\end{figure*}

\begin{figure}
\centering
\includegraphics[width=0.5\textwidth,clip]{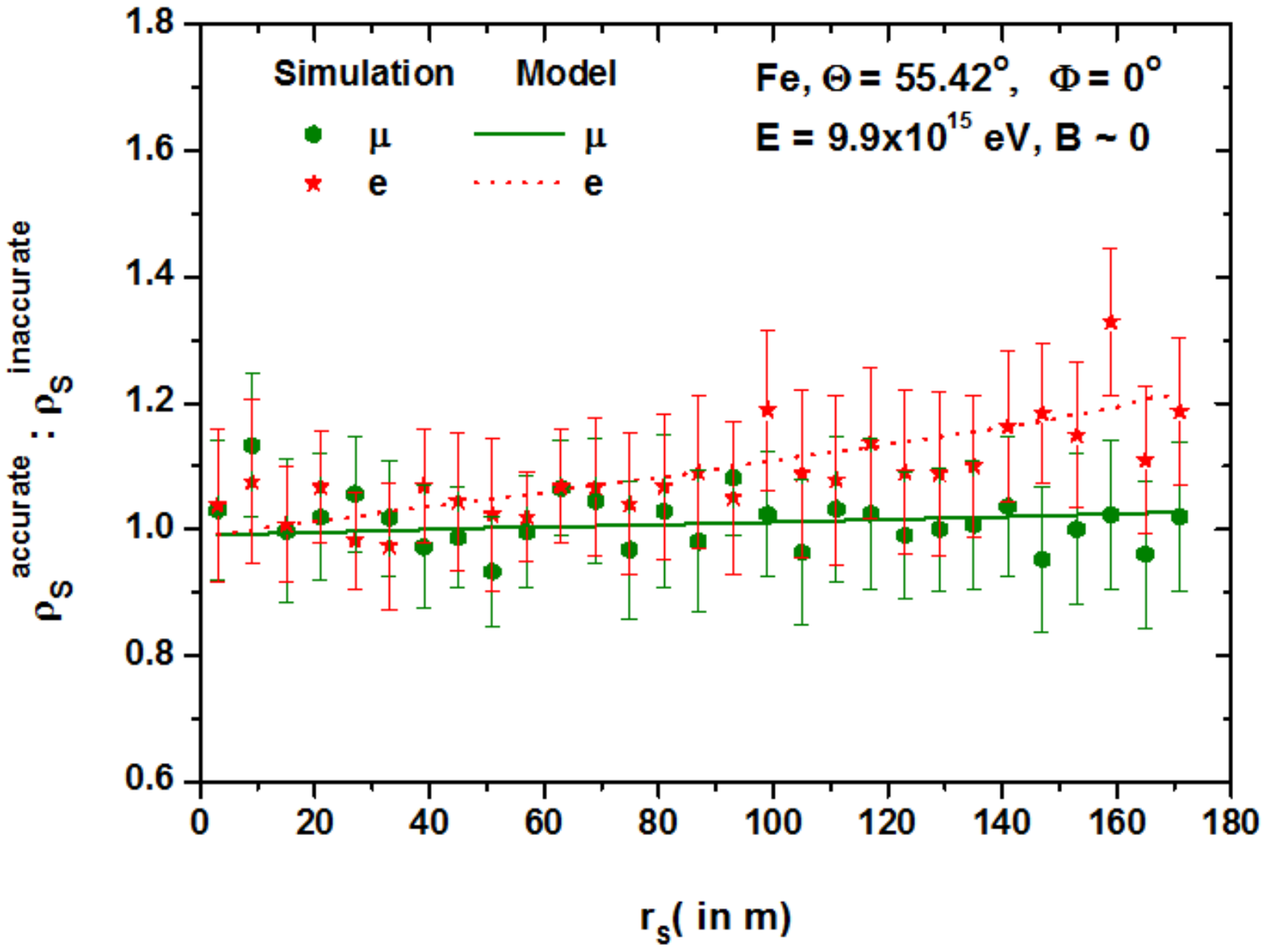}
\caption{The ratios of the accurate (projection+attenuation) to the inaccurate (projection) densities for showers coming from the magnetic north with core distance in the shower front plane. Densities are taken from the core to the late part of showers. The model gives the ratio as, $\rm{e}^{-({\eta x_{g}sin\Theta}}) \approx (1-{\eta \rm{x}_{g}sin\Theta})$ for the late part of the shower, where $\rm{x_{g}}$ is negative. The pair of lines describing the ratios from the model.}
\end{figure} 

A projection method is applied for transforming the position of hit of each muon from the ground plane onto the shower front plane. This makes actually a transformation of $\rho_{\mu}$ or $N_{{\mu}^{\pm}}$ data from the ground plane onto the shower front plane. In Fig. 2, the transformation of a point of hit by a muon in the ground plane with coordinates; ($r_{\rm g}$,$\phi_{\rm g}$ or $\rm{x}_{\rm g}$, $\rm{y}_{\rm g}$) onto the front plane with the new set as; ($r_{\rm s}$,$\beta_{\rm s}$ or $\rm{x}_{\rm s}$, $\rm{y}_{\rm s}$), is described. The relevant transformation relations are as follows,
  
\begin{equation}
r_{\rm{s}} = r_{\rm{g}} \sqrt{1-sin^{2}\Theta cos^{2}(\phi_{\rm{g}}-\Phi)}
\end{equation}

\begin{table*}
\begin{center}
\begin{tabular}
{|l|l|l|l|r|} \hline

 {\rm{Muons}}  & {\rm{r-range (m)}} & $\beta_{s}:150^{\rm o} - 195^{\rm o} (\%)$ & $\beta_{s}:330^{\rm o} - 375^{\rm o} (\%)$  & {\rm {Total - muons}}    
\\ \hline

$\mu^{+}$   & $30-60$   & $34.75\pm{3.53}$   & $6.53\pm{8.91}$   & 2052    \\ 
$\mu^{-}$   & $30-60$   & $6.55\pm{8.43}$   & $34.61\pm{3.60}$   & 2019    \\ \hline      

$\mu^{+}$   & ${\bf 60-90}$   & ${\bf 38.27\pm{4.29}}$   & ${\bf 5.28\pm{9.47}}$   & ${\bf 1660}$    \\ 
$\mu^{-}$   & ${\bf 60-90}$   & ${\bf 5.10\pm{10.34}}$   & ${\bf 38.45\pm{5.09}}$   & ${\bf 1646}$    \\ \hline      

$\mu^{+}$   & $90-120$   & $41.47\pm{5.86}$   & $4.22\pm{14.29}$   & 1201     \\ 
$\mu^{-}$   & $90-120$   & $4.12\pm{13.29}$   & $41.93\pm{5.76}$   & 1192     \\ \hline  							
\end{tabular}
\caption {Selection criteria for best possible muon detection regions for the estimation of the TMBS parameter.  Here, we have used Fe showers with $\rm {E} = 98 - 102$ PeV, $\Theta = 63^{\rm  o} - 68^{\rm o}$ and $\Phi = 47.5^{\rm o} - 57.5^{\rm o}$.  High energy muons with $p_{\mu} = 10^2 - 10^3$ GeV/c are selected for further analysis. Columns 3 and 4 represent the percentages of muon species arriving in those $\beta_{s}$ intervals. Column 5 accounts the total number of muon species arriving in the corresponding $\rm{r-}$ranges. Highlighted $\rm{r-}$range is selected for the analysis.} 
\end{center}
\end{table*}
\begin{table*}
\begin{center}
\begin{tabular}
{|l|l|l|l|l|l|l|r|} \hline

 {\rm{Muons}}& {\rm{E (GeV)}}& $\rm{p}_\mu$ (GeV/c)& $\beta_{s}:150^{\rm o} - 195^{\rm o}$ ($\%$)& $\beta_{s}:330^{\rm o} - 375^{\rm o}$ ($\%$)& $\rm{p}_\mu$ (GeV/c)& $\beta_{s}:150^{\rm o}-195^{\rm o}$ ($\%$)& $\beta_{s}:330^{\rm o}-375^{\rm o}$ ($\%$)\\ \hline

$\mu^{+}$& $10^6$& $1 - 10^2$& $24.26\pm{39.21}$& $10.21\pm{58.45}$& ${\bf 10^2 - 10^3}$& ${\bf 43.90\pm{26.22}}$& $4.32\pm{85.14}$ \\ 
$\mu^{-}$& $10^6$& $1 - 10^2$& $11.42\pm{55.64}$& $25.11\pm{38.90}$& ${\bf 10^2 - 10^3}$& $4.06\pm{88.40}$& ${\bf 43.93\pm{26.83}}$ \\ \hline  
$\mu^{+}$& $10^7$& $1 - 10^2$& $25.17\pm{10.10}$& $11.41\pm{17.11}$& ${\bf 10^2 - 10^3}$& ${\bf 41.73\pm{8.21}}$& $4.46\pm{36.94}$ \\
$\mu^{-}$& $10^7$& $1 - 10^2$& $10.55\pm{20.26}$& $24.70\pm{11.88}$& ${\bf 10^2 - 10^3}$& $4.70\pm{27.95}$& ${\bf 39.88\pm{12.32}}$ \\ \hline
$\mu^{+}$& $10^8$& $1 - 10^2$& $24.97\pm{6.39}$& $10.48\pm{11.81}$ & ${\bf 10^2 - 10^3}$& ${\bf 38.27\pm{4.29}}$& $5.28\pm{9.47}$ \\
$\mu^{-}$& $10^8$& $1 - 10^2$& $11.29\pm{10.08}$ & $24.57\pm{6.78}$& ${\bf 10^2 - 10^3}$& $5.10\pm{10.34}$& ${\bf 38.45\pm{5.09}}$ \\ \hline   							
\end{tabular}
\caption {Analysis showing an implementation of the selection of muons momenta in selected detection regions obtained from Table III. Here, we have used Fe showers with $E = 1-3$, $8-12$, and $98-102$ PeV, and $\Theta = 63^{\rm  o} - 68^{\rm o}$, and $\Phi = 47.5^{\rm o} - 57.5^{\rm o}$. Columns 4, 5, 7 and 8 represent the percentages of muon species arriving in those $\beta_{s}$ intervals. Highlighted $\rm{p_{\mu}}-$range is selected for the analysis.} 
\end{center}
\end{table*}

or, equivalently;

\begin{equation}
x_{\rm{s}} = x_{\rm{g}}cos\Theta = r_{\rm{g}}cos(\phi_{\rm{g}}-\Phi)cos\Theta
\end{equation} 
\begin{equation} 
y_{\rm{s}} = y_{\rm{g}} = r_{\rm{g}}sin(\phi_{\rm{g}}-\Phi)
\end{equation}

The muon density in the shower front plane ($\rho_{\rm s}(r_{\rm s},\beta_{\rm s}$)) can be obtained from the muon density measured in the ground plane ($\rho_{\rm g}(r_{\rm g},\phi_{\rm g}$)) by a simple geometrical transformation, and is given by         
\begin{equation}
\rho_{\rm{s}}^{\rm{inaccur.}}(r_{\rm{s}},\beta_{\rm{s}}) = \frac{\rho_{\rm{g}}(r_{\rm{g}},\phi_{\rm{g}})}{cos{\Theta}}.
\end{equation} 

The $\rho_{\rm s}^{\rm{inaccur.}}(r_{\rm s},\beta_{\rm s}$) obtained from the above, is inaccurate as because they do not involve the attenuation of muons in the region between the two planes. An accurate measure for the muon density ($\rho_{\rm s}^{\rm{accur.}}(r_{\rm s},\beta_{\rm s}$)) in the shower plane [18], taking account the geometrical projection of $\rho_{\rm g}(r_{\rm g},\phi_{\rm g}$) onto the shower front plane with attenuation correction, is done through 
\begin{equation}
\rho_{\rm{s}}^{\rm{accur.}}(r_{\rm{s}},\beta_{\rm{s}}) = \frac{\rho_{\rm{g}}(r_{\rm{g}},\phi_{\rm{g}})}{cos{\Theta}}~e^{{\pm}({\eta x_{g}sin\Theta})},
\end{equation}   
where $\eta$ = $\frac{\Delta{X}}{\lambda}$ gives a measure for the attenuation length in $\rm m^{-1}$. The attenuation length $\lambda$ for electrons at the KASCADE level is about 190 g cm$^{-2}$. For the muonic component, $\lambda$ gets a value close to 900 g cm$^{-2}$. In our calculation, near the surface of the Earth the increase $\Delta{X}$ is approximately equal to 0.15 g cm$^{-2}$ for every meter traveled through the atmosphere. With the above substitution $\eta$ takes $\sim \frac{0.15}{190}$ and $\sim \frac{0.15}{900}$ m$^{-1}$ respectively for $\rm e^{\pm}$ and ${\mu}^{\pm}$ at the KASCADE level [19,20]. For a cylindrical EAS profile, we have to substitute $\pm{x_{\rm g} sin\Theta}$ ($-$ and $+$ signs account the attenuation of the late and early parts of the EAS) for the extra path length between the planes.  Equations 4 and 5 are true for the showers coming from the magnetic north direction i.e. $\Phi = 0^{\rm o}$ in \emph{CORSIKA} plane. For an arbitrary direction ($\Phi$), one should use $\phi_{\rm g}-{\Phi}$ for $\phi_{\rm g}$ in those equations. We will now display the ratio between the accurate and inaccurate muon densities against the radial distance (${r_{\rm s}}$) for simulated muon densities when ${\rm B}$ is $\sim 10^{-5}\times {\rm B}_{\rm{KAS.}} \sim 0$. From the geometry of Fig. 2, it also follows  

 \begin{equation}
{r_{\rm s}} = (x_{\rm g}^{2}cos^{2}\Theta + y_{\rm g}^{2})^{\frac{1}{2}}.
\end{equation}

The same procedure can be applied to simulated electron data. At ${\rm B} \sim 0$, the ratio between the accurate and inaccurate muon or electron densities in the late part of the shower plane are shown in the Fig. 3 with ${r_{\rm s}}$. The scattered points represent simulated data while the lines are obtained directly from the equations 4 and 5 [18]. The figure indicates that the muon densities in the shower front plane with or without the attenuation corrections are nearly the same, and are negligibly smaller than what is contributed by the charged EM component of a shower. Hence, attenuation of muons in the region between the two planes can be ignored judiciously in the data analysis.

From the viewpoint of the experimental feasibility of the present method, this analysis selects muons that fulfill two selection criteria simultaneously involving muon detection area and muon energies or momenta. High-energy muons get going from the upper part of the atmosphere and hence experience the GMF for a longer duration as they are approaching the ground level. This is consistent with the expectations since the high energy muons will last long against attenuation in highly inclined showers.

The TMBS and MTMBS parameters are expected to be affected by the energy of incoming muons. This work has identified some suitable conditions for the detection and the measurement of these parameters with reasonable statistics of muons. Therefore a best compromise among the muon detector size, muon energy thresholds, and the $N_{\mu}^{\rm{tr.}}$ must be achieved. The parameter $N_{\mu}^{\rm{tr.}}$ accounts the total number of muons in a region between $60$ m and $90$ m core distances with $10^{\rm o}$ or $15^{\rm o}$ polar angle bin made by two diagonal lines passing through the EAS core. The noticeable effects of the GMF are emphasized in the case of highly inclined showers with high momentum muons ($10^2 - 10^3$ GeV/c). 

A practical realization of the present approach requires a pair of muon detectors covering two finite regions in diagonally opposed positions, will be shown in the final Fig. 10 in conjunction with an EAS array containing several scintillation detectors. It is noteworthy to state that the azimuthal modulation of shower events around the two aforesaid opposed azimuthal windows arises due to geomagnetic effect. For example, at the KASCADE location, we have observed such azimuthal modulations of shower events around $47.5^{\rm o}-57.5^{\rm o}$ and $240^{\rm o}-250^{\rm o}$ in our simulation with full azimuthal range $\Phi = 0^{\rm o}-360^{\rm o}$. This work takes only these events for further analysis at the KASACADE site.

The work also foresees the employment of a pair of muon detectors into the array with some reasonable sizes from the point of view of their construction cost. We have used different combinations of momentum thresholds and muon detection areas as trials so that a better option may be surfaced, which will deliver an optimal asymmetry between $\mu^+$ and $\mu^-$ particles. It has been seen that the low momenta muons (below $\sim 10^2$ GeV/c) offer nearly symmetrical distribution in the $\rm{X-Y}$ shower plane, whereas muons with momenta falling in the range $10^2 - 10^3$ GeV/c manifest a better polar asymmetry in $\mu^+$ and $\mu^-$ numbers in the annular region between $60$ m and $90$ m from the EAS core. Table III and Table IV give a clear view on muon selections in accordance with the above requirement. The highlighted figures in both the tables correspond to the best combinations between muons thresholds and their detection areas, for which an optimum local contrast in the abundance of $\mu^{+}$ and $\mu^{-}$ could be achieved. The relative errors associated with these percentages are provided, in order to get an indication of the significance of these percentage figures.  Non-highlighted figures in those tables account lower percentage of $\mu^{+}$ and $\mu^{-}$ abundances at two detector locations out of a relatively higher total in the whole annular region. In Table III, higher percentages of $\mu^{+}$ and $\mu^{-}$ are found in a relatively lower total in the annular region between 90 m and 120 m from the EAS core. The error on these percentages also justify the above facts. Uncertainties in the Table IV in particular, decrease with the increase of primary energy simply because of increasing muon number statistics. Other than highlighted combinations, are found unsuitable for estimating TMBS and MTMBS with some anticipated uncertainties. In the upcoming sections, we shall consider only these selected muons having $\rm{p}_{\mu} = 10^2 - 10^3$ GeV/c and $\rm{r} = 60 - 90$ m irrespective of geographical sites. 

\section{Results and discussions}
\subsection{Studying asymmetry of high energy muons in EAS}

\begin{figure*}[ht]
\subfigure
{\includegraphics[scale=0.25]{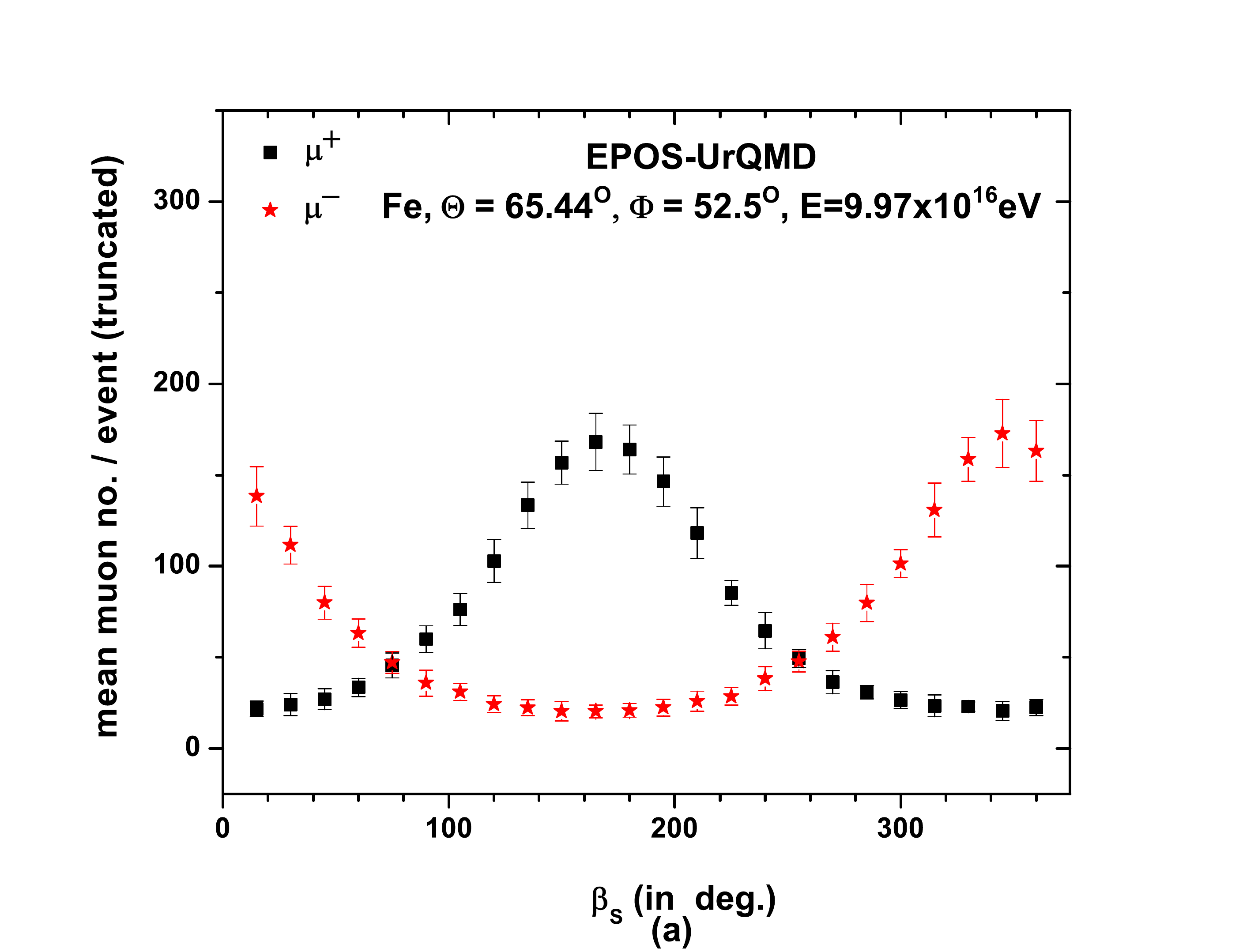}}
\subfigure 
{\includegraphics[scale=0.25]{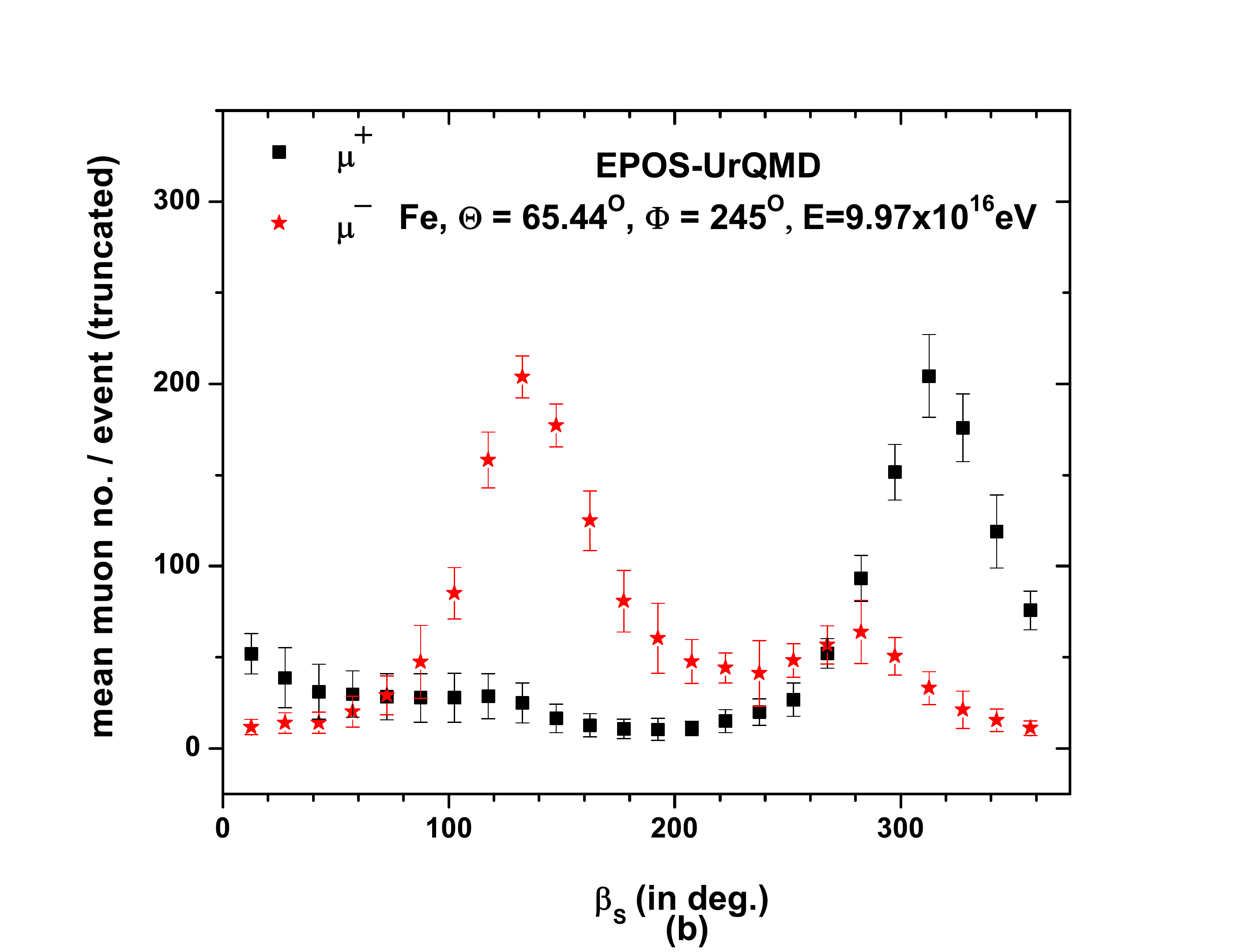}}
\\
\subfigure
{\includegraphics[scale=0.25]{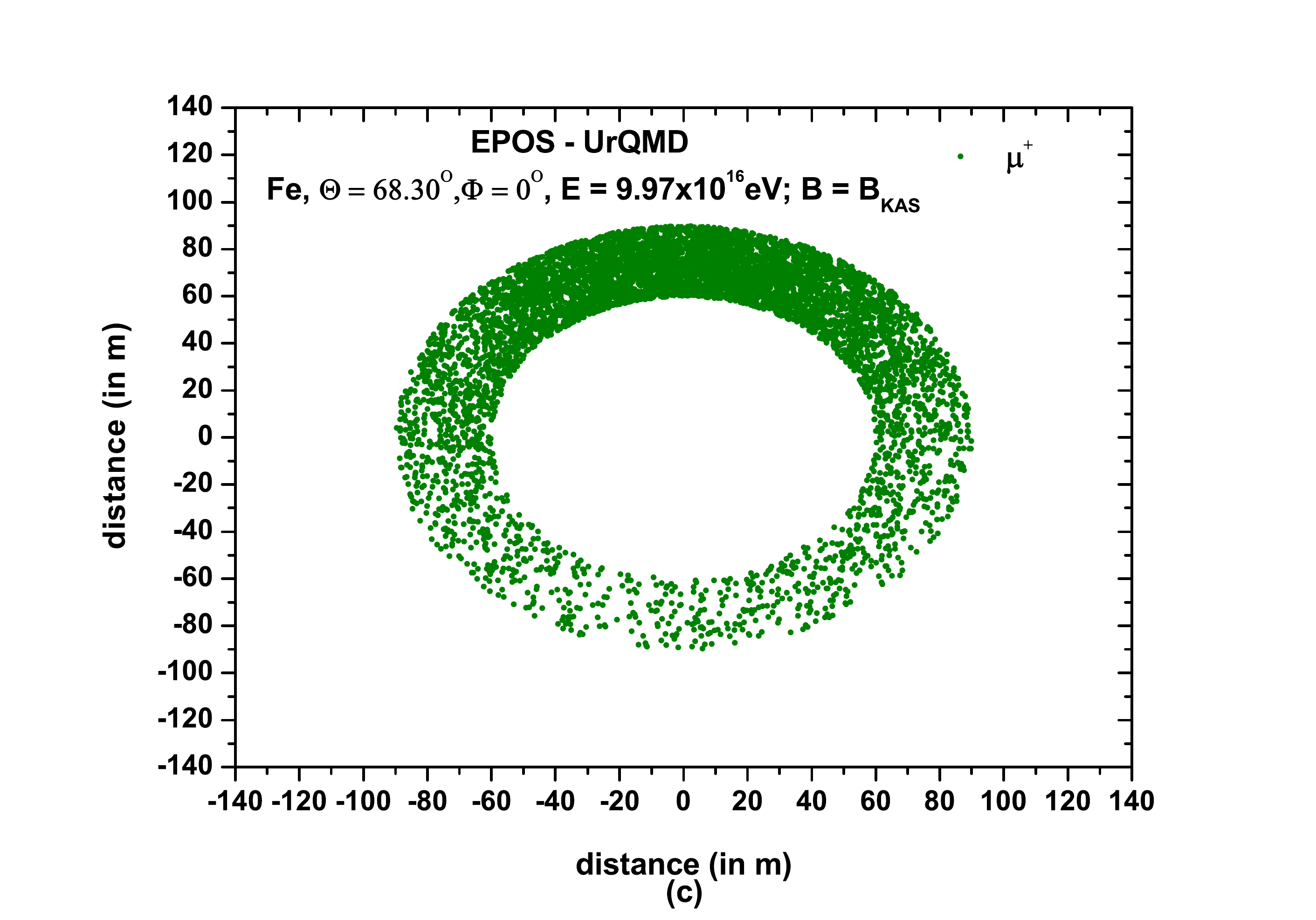}}
\subfigure 
{\includegraphics[scale=0.25]{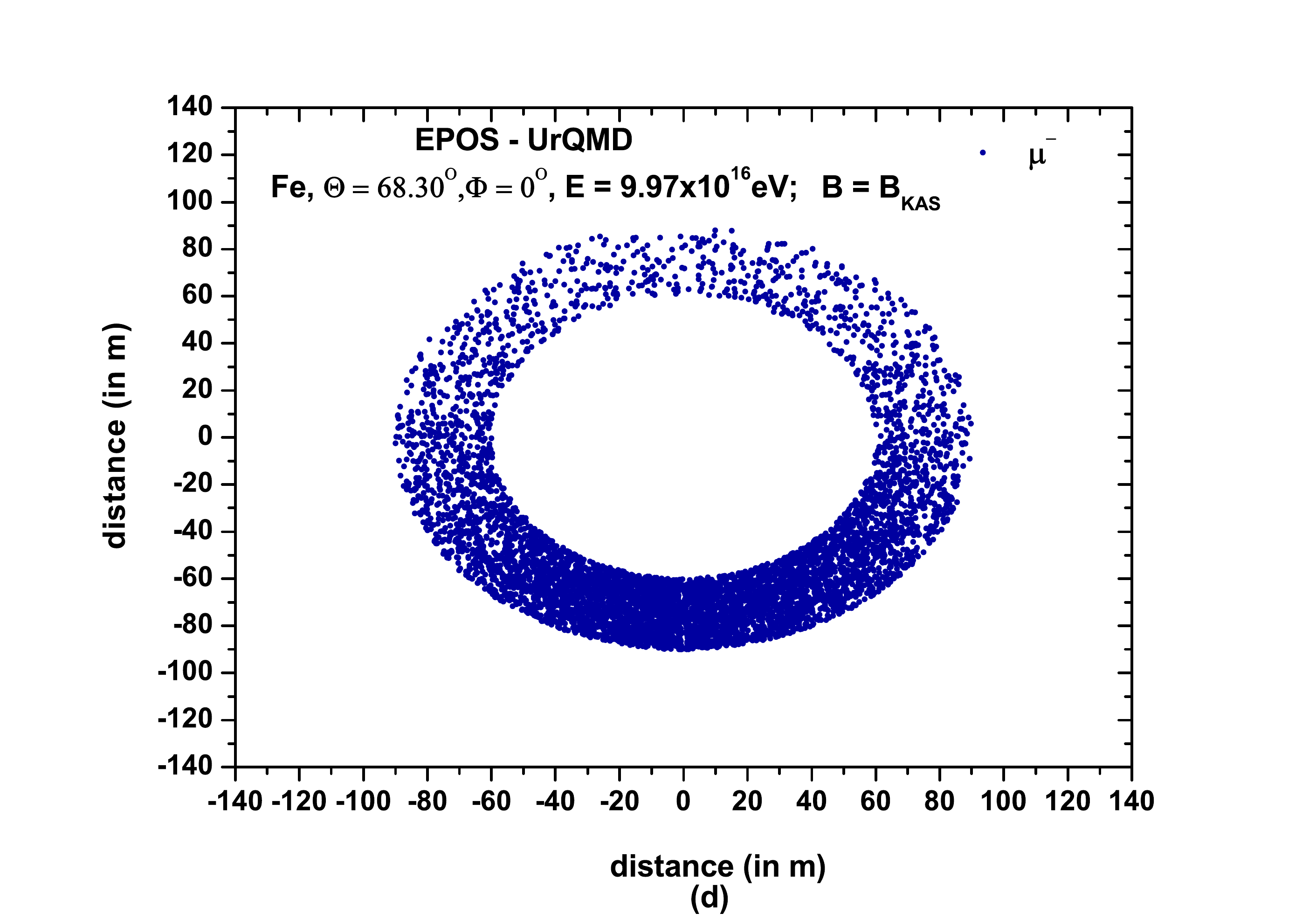}}
\\
\subfigure
{\includegraphics[scale=0.25]{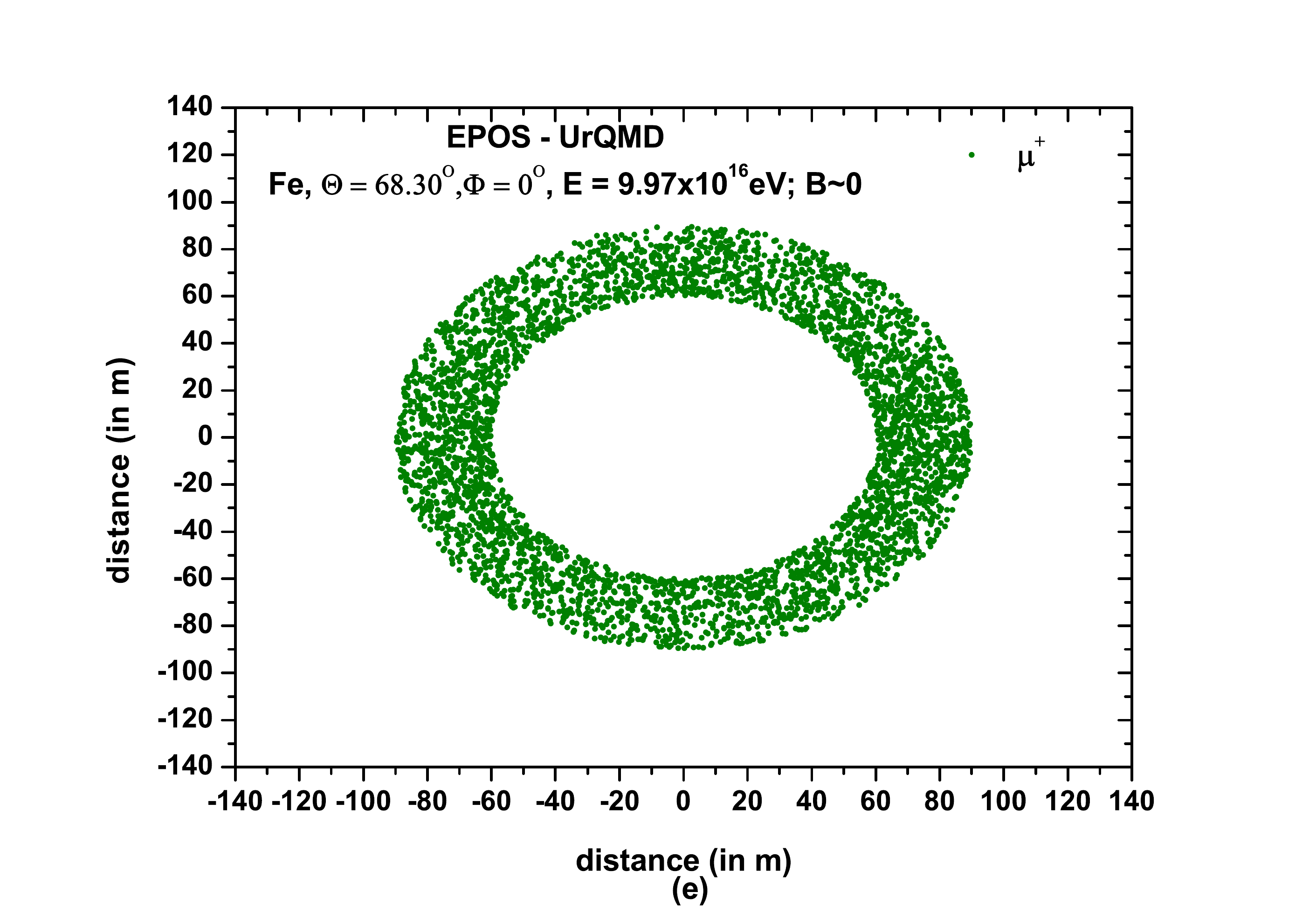}}
\subfigure 
{\includegraphics[scale=0.25]{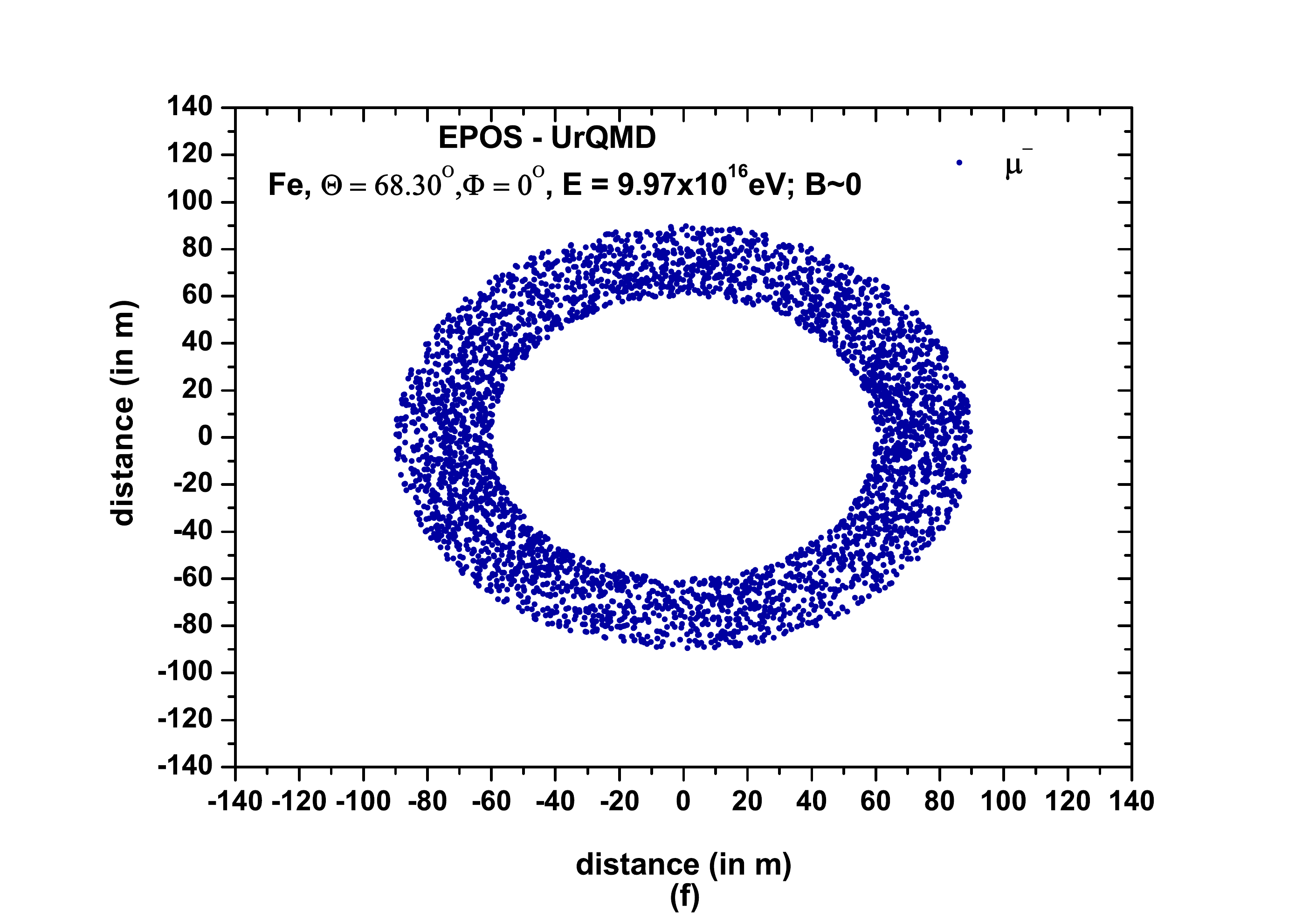}} 

\caption{In Fig. a and Fig. b, we have used our two selected $\Phi$ ranges but keeping the mean $\Theta$ at $65.44^{\rm o}$ for the mean polar variations of $\mu^{+}$ and $\mu^{-}$ for iron primary. The mean polar variations of $\mu^{+}$ and $\mu^{-}$ are shown in scattered figures c and d when $\rm{B}\sim \rm{B}_{\rm{KAS.}}$, and also in figures e and f when $\rm{B}\sim 0$. For Fig. a and b, the x-label represents the polar angle; $\beta_{\rm s}$ while for figures c, d, e, and f, the x-label is a distance. Data from the shower plane are used only.} 
\end{figure*}  

We have estimated total number of $\mu^{+}$ and $\mu^{-}$ over a region between the arc of radii $60$ m and $90$ m, with a central angle of amount $\sim 15^{\rm o}$ on the shower plane when GMF is active (taking $\rm{p}_{\mu} = 10^2 - 10^3$ GeV/c). The Fig. 4a shows polar asymmetries of $\mu^{+}$ and $\mu^{-}$ for $\langle \Theta\rangle = 65.44^{\rm o}$ and the selected $\langle \Phi\rangle = 52.5^{\rm o}$ at KASCADE level. We have repeated the same study to obtain the Fig. 4b but for showers having $\langle \Phi\rangle = 245^{\rm o}$. It is noticed from these figures that for $\langle \Phi\rangle = 52.5^{\rm o}$, the modulations of muon species are seen around $\beta_{\rm s} \sim 165^{\rm o}$ and $\sim 345^{\rm o}$, while for $\langle \Phi\rangle = 245^{\rm o}$, around $\beta_{\rm s} \sim 132^{\rm o}$ and $\sim 312^{\rm o}$ indicate the prominent modulation regions. These curves would therefore validate our selections of polar angle ranges that have been worked out in Table III and Table IV. However, these selected oppositely aligned polar angle ranges for the installation of muon detectors will definitely vary from one experimental location to other. Because, the influence of the GMF on the azimuthal distribution of EAS muons/particles of a location depends on the direction of the incoming CR shower, thereby introducing a modulation in the azimuthal distribution of CRs. It is clear that the path of an EAS muon is modified by the Lorentz force in the bending plane. A given muon with a momentum vector $\stackrel{\rightarrow}{p}_{\mu}$, the Lorentz force acting on it depends on the magnetic field $\stackrel{\rightarrow}{B}$ and its angle with $\stackrel{\rightarrow}{p}_{\mu}$. This angle is expected to be a function of shower arrival direction (i.e. $\Theta$ and $\Phi$). For a clear visibility of asymmetries developed in $\mu^{+}$ and $\mu^{-}$ distributions, we have added scattered plots through Fig. 4c, 4d, 4e and 4f, for showers arriving from the very restrictive magnetic north direction.

\begin{figure}
\centering
\includegraphics[width=0.5\textwidth,clip]{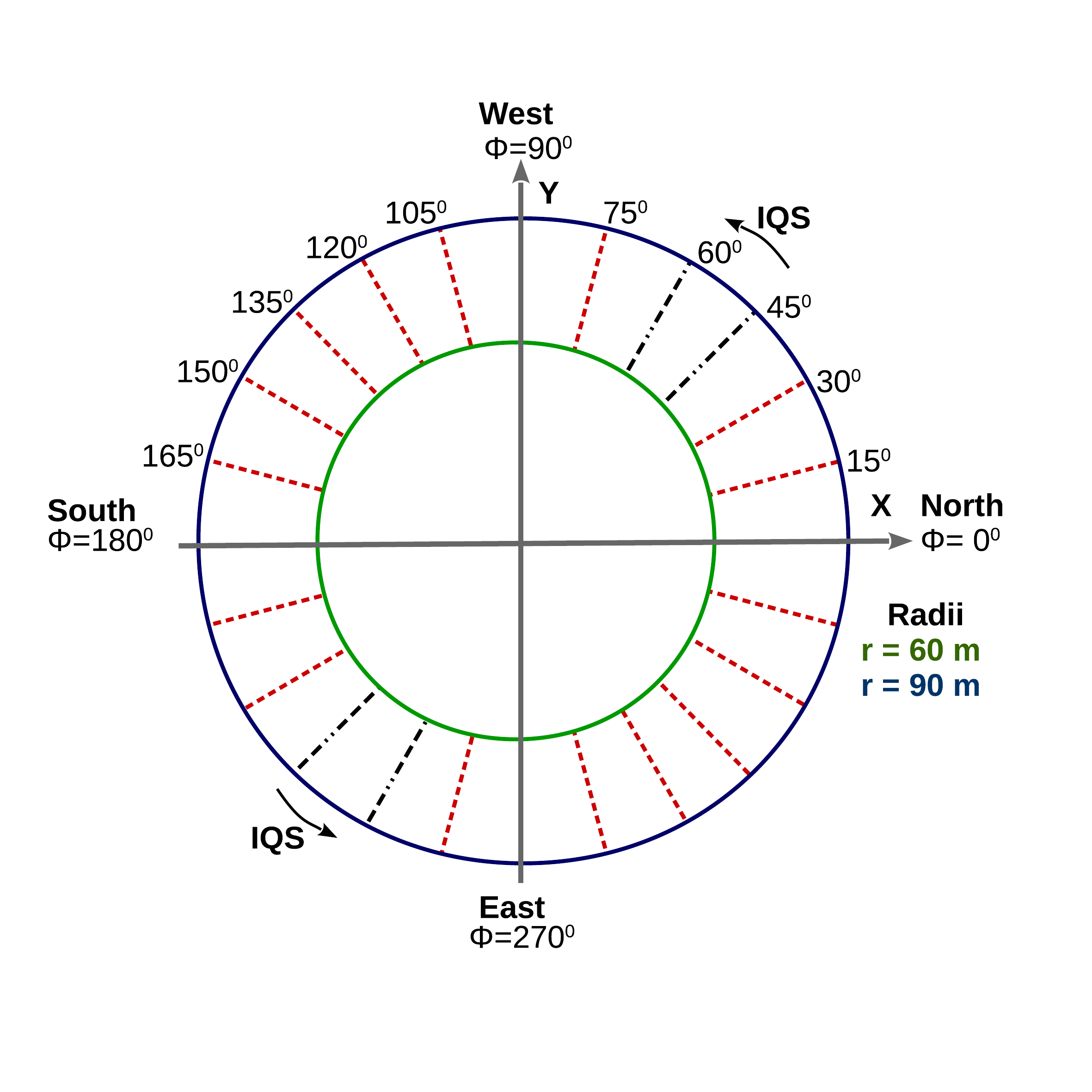} \hfill 
\caption{Scanning of $\mu^{+}$ and $\mu^{-}$ particles' positions by rotating IQS in anti-clockwise sense from $0^{\rm o}$ to $180^{\rm o}$.}
\end{figure} 

\subsection{The transverse muon barycenter separation}

\begin{figure}
\centering
\includegraphics[width=0.5\textwidth,clip]{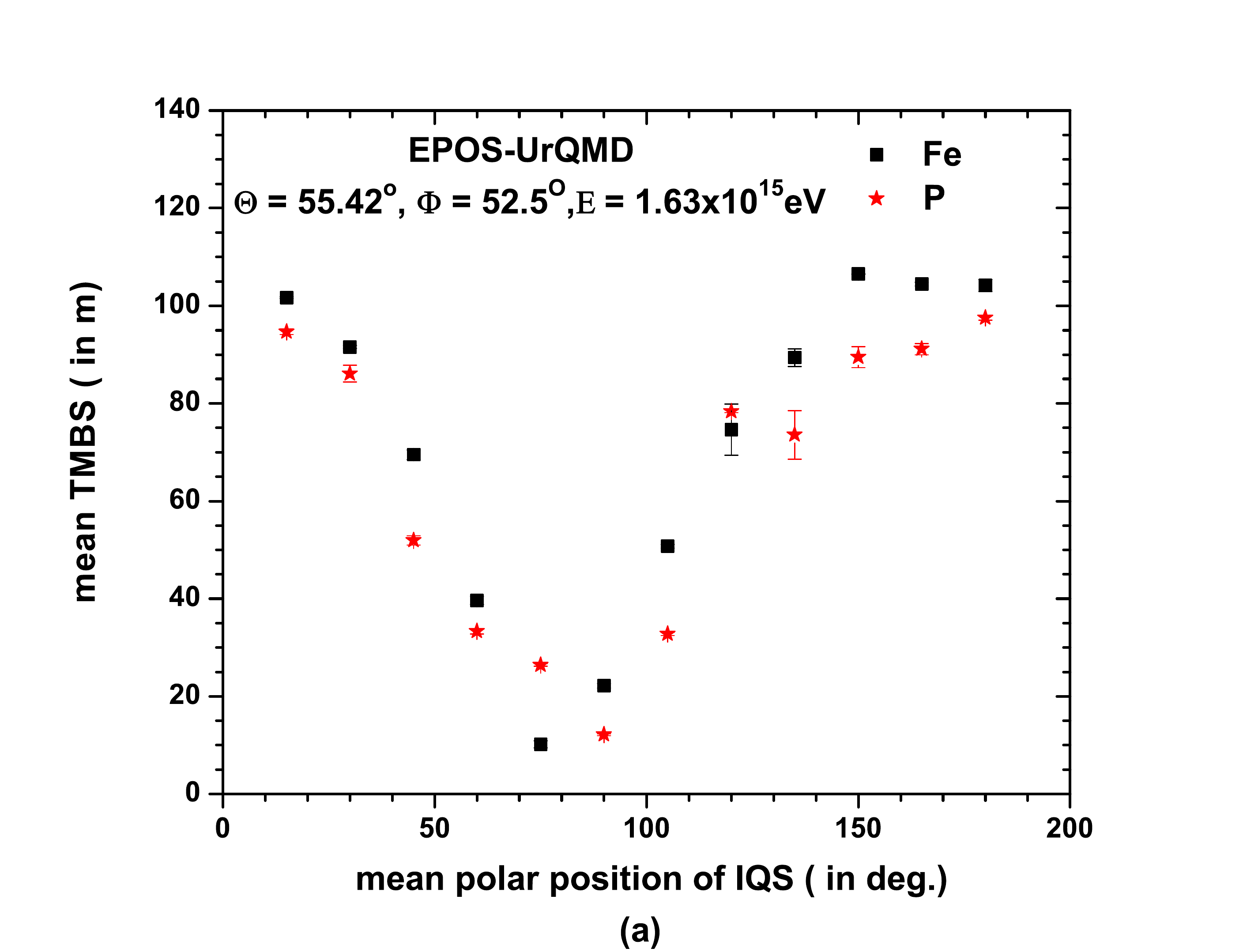} \hfill 
\includegraphics[width=0.5\textwidth,clip]{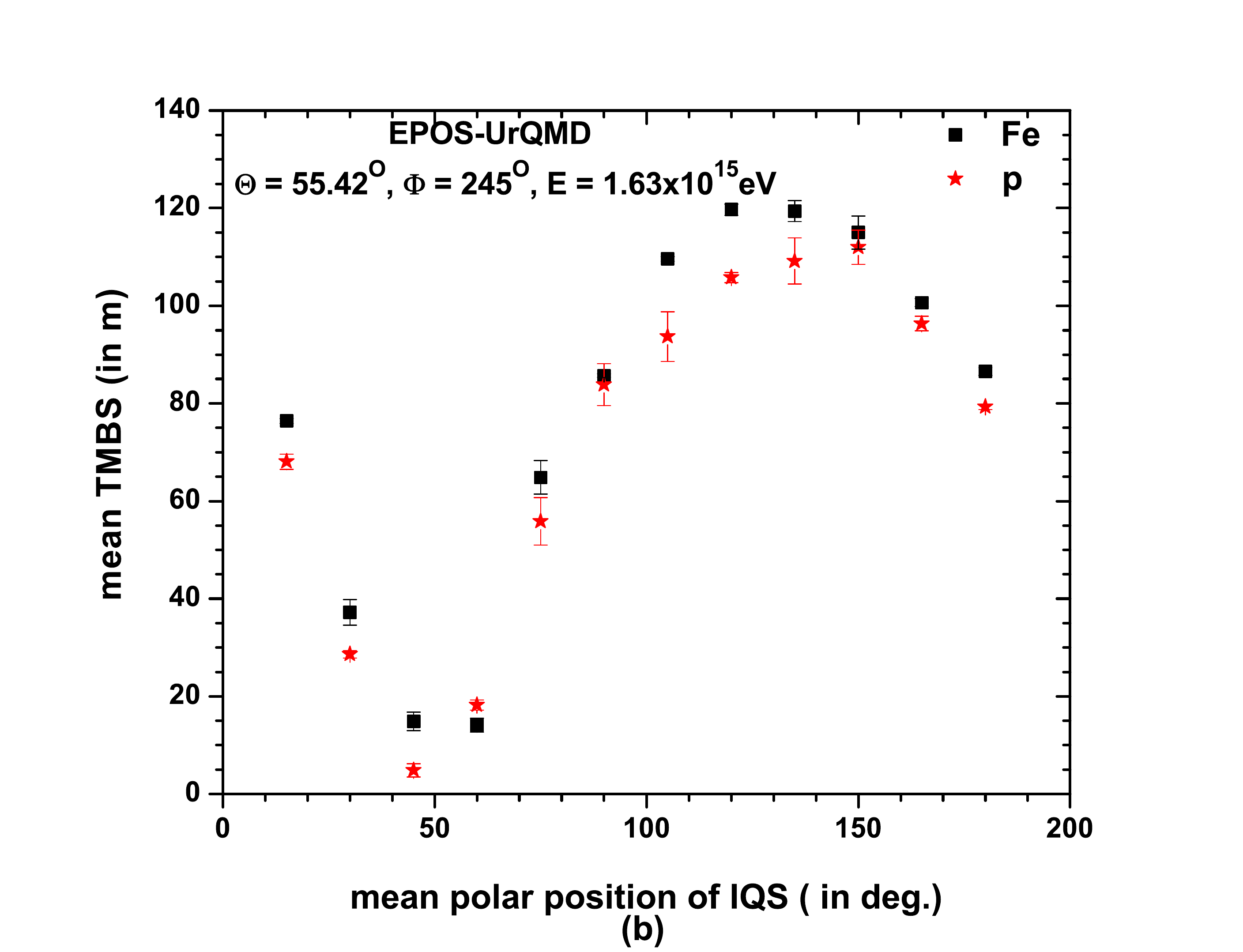} \hfill
\caption{Polar variation of the mean TMBS for p and Fe showers arriving from two mean arbitrary directions.}
\end{figure}

We are now in a position to estimate the coordinates of barycenters of $\mu^{+}$ and $\mu^{-}$ in the shower plane for each shower. The linear distance between the barycenters of $\mu^{+}$ and $\mu^{-}$ from two opposed regions in a hypothetical interior quadrant sector (IQS) is then estimated. This distance is just equal to our desired TMBS parameter in the work. To estimate the parameter at multi-polar positions, an operation is introduced that executes a rotation either clockwise or anti-clockwise sense of the IQS for estimating the $\mu^{+}$ and $\mu^{-}$ positions. An IQS represents a region in the interior between two circles enclosed by a pair of arcs on opposite sides and a pair of diagonally aligned straight lines passing through the EAS core making a central angle of $\sim 10^{o}/15^{\rm o}$. The IQS involves two diagonally opposed annular regions between the core distances $60$ m and $90$ m, which is shown in the Fig. 5.

\begin{figure*}[ht]
\subfigure
{\includegraphics[scale=0.25]{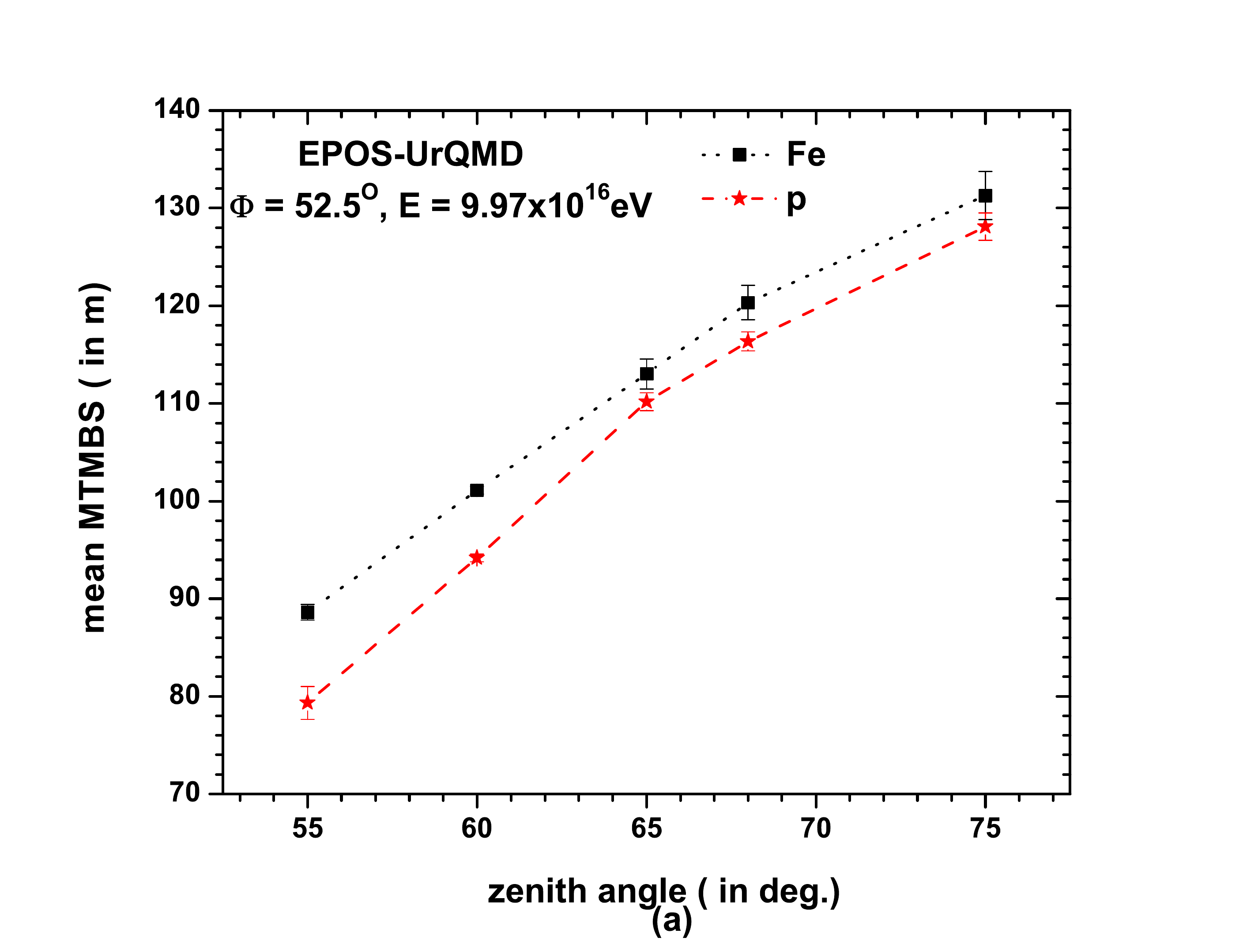}}
\subfigure 
{\includegraphics[scale=0.25]{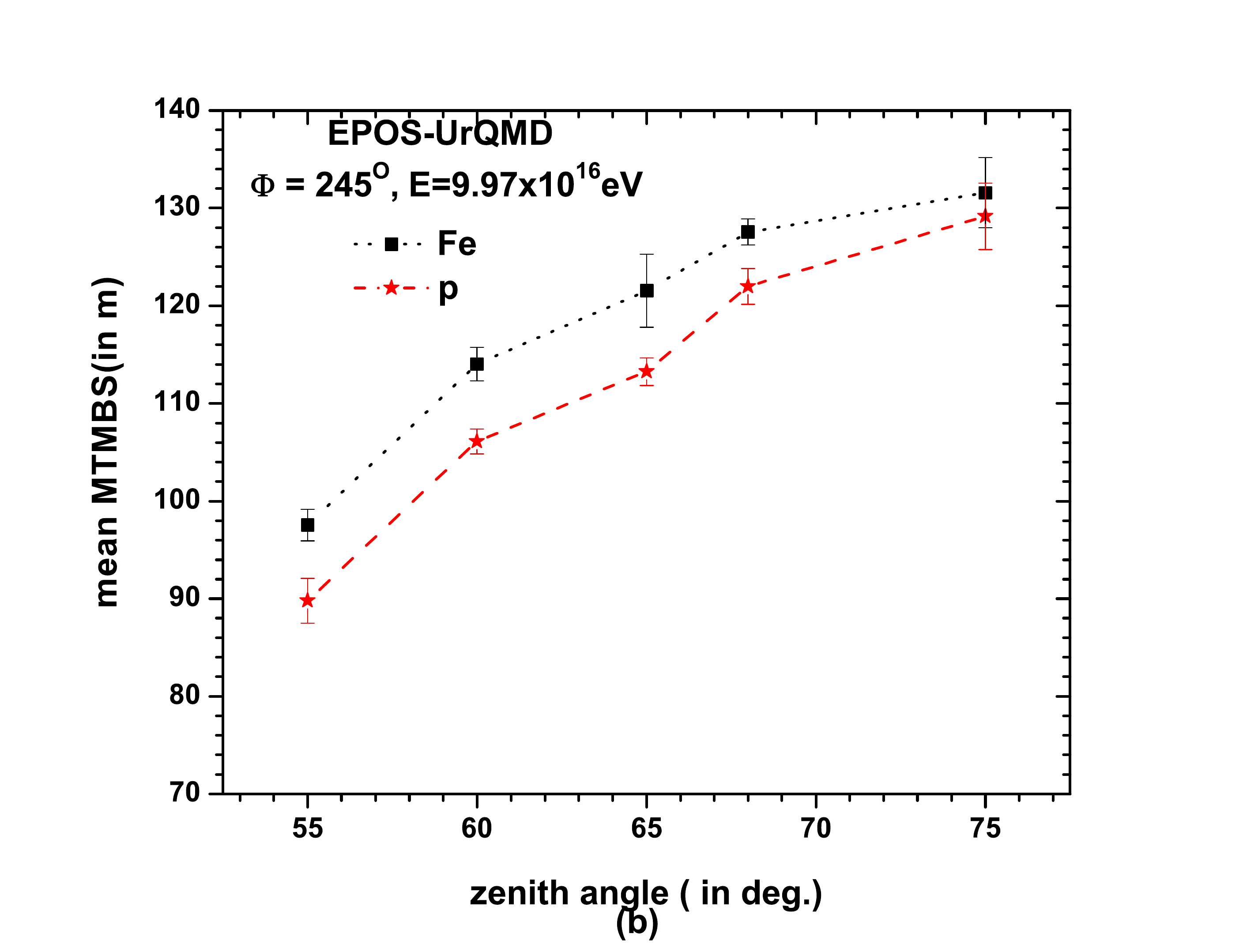}}
\\
\subfigure
{\includegraphics[scale=0.25]{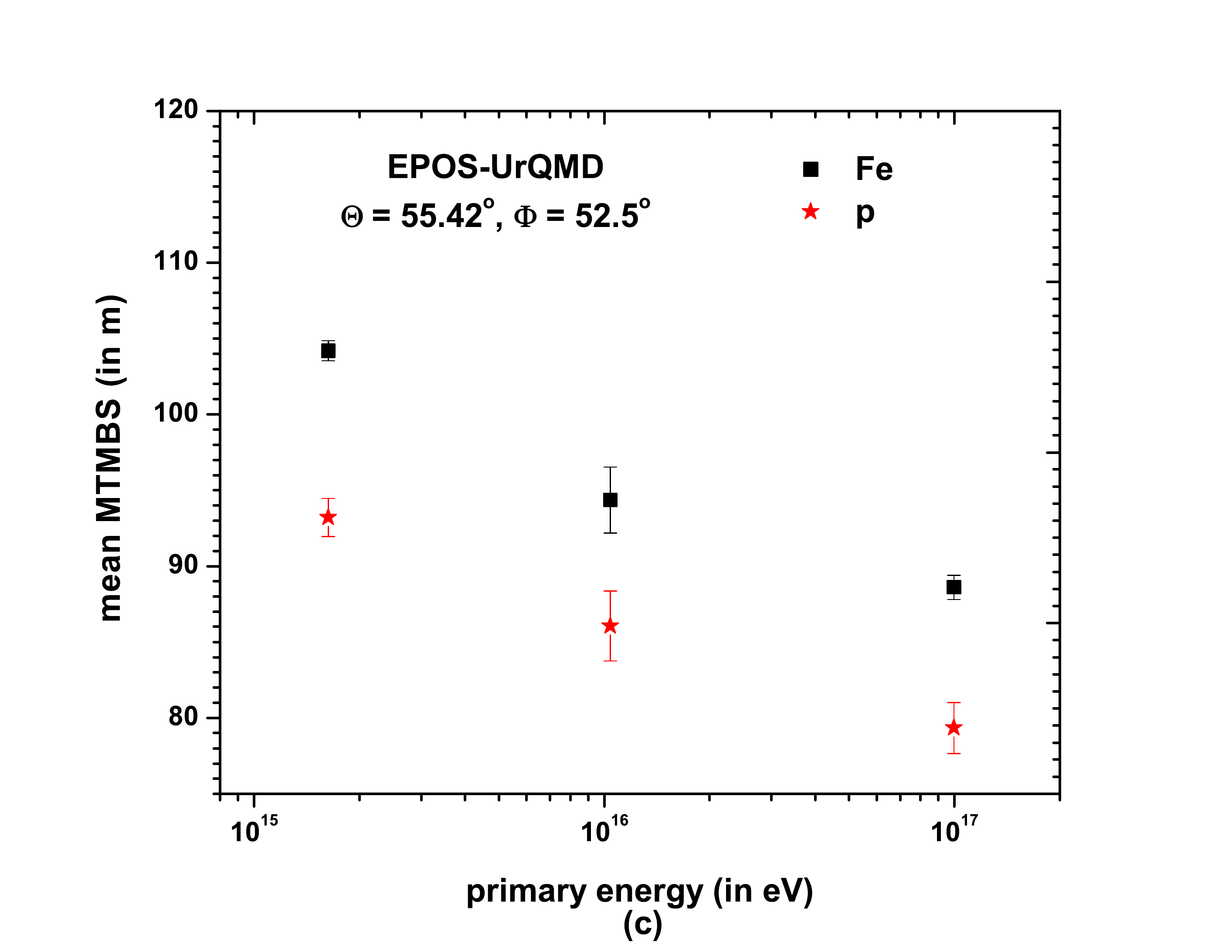}}
\subfigure 
{\includegraphics[scale=0.25]{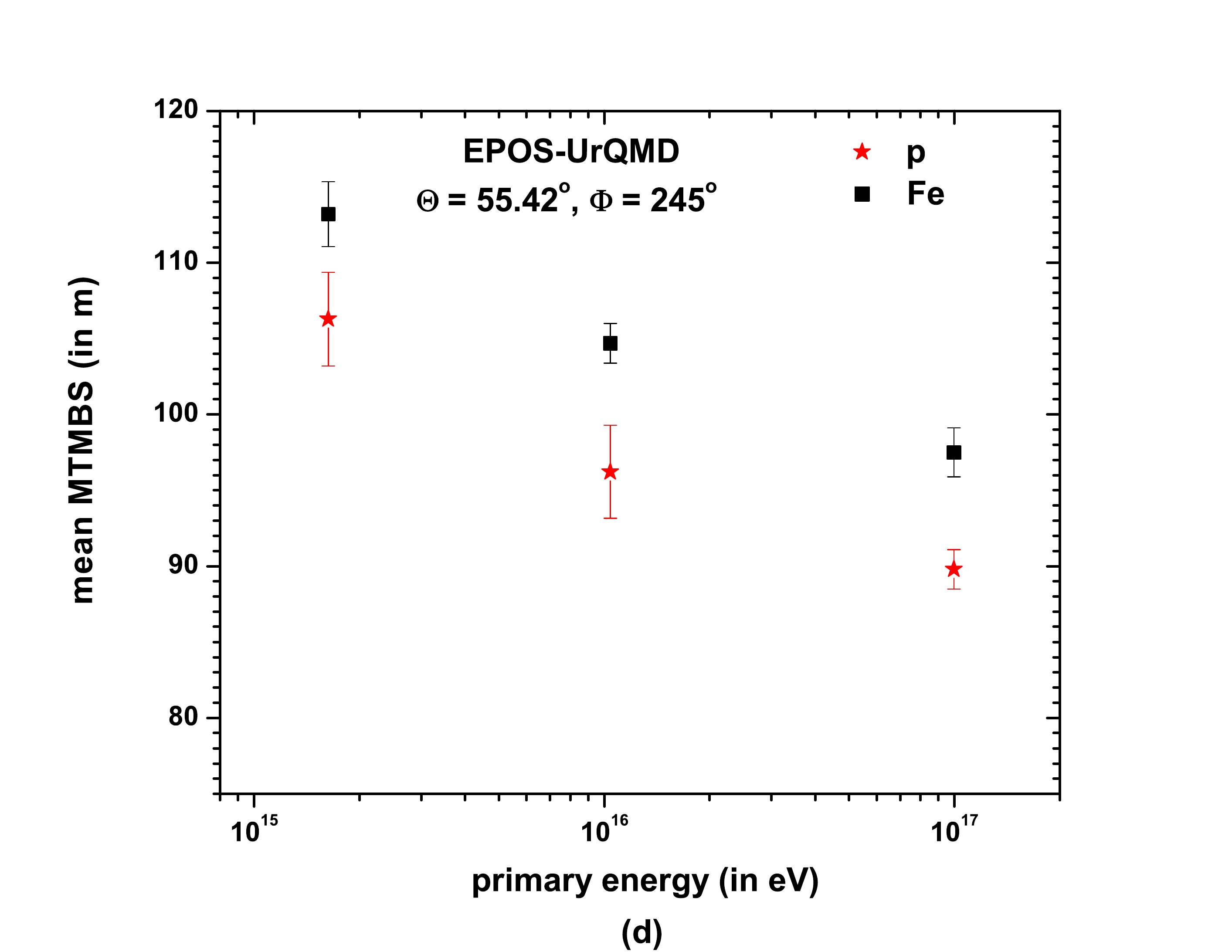}}

\caption{Dependence of the mean MTMBS parameter on zenith angle and primary energy of a shower. The lines are only a guide for the eye in figures a and b.} 
\end{figure*} 

The TMBS is expected to vary with the rotation of IQS. To show that we have used p and Fe initiated EASs arriving from average directions, $\langle \Phi\rangle = 52.5^{\rm o}$ and $\langle \Phi\rangle = 245^{\rm o}$ with the zenith angle range $53^{\rm o} - 58^{\rm o}$. In Fig. 6a and Fig. 6b, a clear contrast of these variations are presented for showers arriving through the azimuthal angle ranges; $47.5^{\rm o} - 57.5^{\rm o}$ and $240^{\rm o} - 250^{\rm o}$ at $\langle \Theta \rangle = 55.42^{\rm o}$. It is expected that $\mu^{+}$ and $\mu^{-}$ particles concentrate mostly around the polar angles $\sim 165^{\rm o}$ and $\sim 345^{\rm o}$ by experiencing the GMF effects, (peaks in the Fig. 4a) when $\langle \Phi\rangle \sim 52.5^{\rm o}$. For $\langle \Phi\rangle \sim 245^{\rm o}$, the Fig. 4b shows these peaks around polar angles $\sim 132^{\rm o}$ and $\sim 312^{\rm o}$ respectively. The TMBS parameter accordingly takes higher values corresponding to the orientation of the IQS through $\sim 165^{\rm o} - 345^{\rm o}$, which is reflected in the Fig. 6a for $\langle \Phi\rangle \sim 52.5^{\rm o}$. For $\langle \Phi\rangle = 245^{\rm o}$, the TMBS parameter versus rotation of the IQS is shown in the Fig. 6b. In this case the maximum value of the TMBS parameter is found along the $\sim 132^{\rm o} - 312^{\rm o}$ orientation of the IQS. 

The TMBS parameter, following the IQS position with highest concentrations of $\mu^{+}$ and $\mu^{-}$ in a diagonally opposite regions, has been estimated finally. This is called the maximum TMBS (i.e. MTMBS) parameter, and such a parameter will be used extensively to correlate Earth's geomagnetic activity with high energy muons. For a given sample of showers, e.g. $E= 1-3$ PeV, $\Theta = 53^{o} - 58^{o}$, $\Phi=47.5^{o} - 57.5^{o}$, we have obtained the average behaviour of the TMBS as shown in figure 6a and similarly for the figure 6b and so on. As the maximum value of the TMBS varies over a small spread from shower to shower for the same sample of showers, it is therefore justified to estimate the mean maximum TMBS about the center of the spread. For each shower, we have an average value corresponding to about 5 TMBS values estimated for 5 diagonal positions of the IQS centered about $172.5^{o}-352.5^{o}$ position. For all the showers from the sample, the mean MTMBS parameter can then be obtained. The variation of the mean MTMBS against $\Theta$ is shown in Fig. 7a and Fig. 7b for showers induced both by p and Fe, with $\langle \Phi\rangle \simeq 52.5^{\rm o}$ and $\simeq 245^{\rm o}$ respectively at $\langle E \rangle \approx 9.97 \times 10^{16}$ eV. Through the Fig. 7c and Fig. 7d, the CR mass sensitivity of the MTMBS parameter is demonstrated.

\subsection{Correlations of the GMF components with MTMBS}

Now, we shall investigate the impact of the GMF components on the MTMBS parameter. Proton and iron initiated showers with $1$ PeV primary energy have been used. The steering file in the \emph{CORSIKA} code is made ready for two arbitrary conditions: (i) $\Theta = 50^{\rm o}$ and $\phi = 0^{\rm o}$, (ii) $\Theta = 60^{\rm o}$ and $\phi = 55^{\rm o}$. For the locations in Table I and II, the geographical latitude and longitude, the total magnetic field ($\rm B$) and its horizontal ($\rm B_{H}$), and vertical ($\rm B_{V}$) components are known from the scientific research organization; NOAA [17]. By the application of the present technique the dependence of the mean MTMBS parameter on $\rm B$ and its components $\rm B_{H}$ and $\rm B_{V}$ can be studied for different locations. 
\begin{figure*}[ht]
\subfigure
{\includegraphics[scale=0.25]{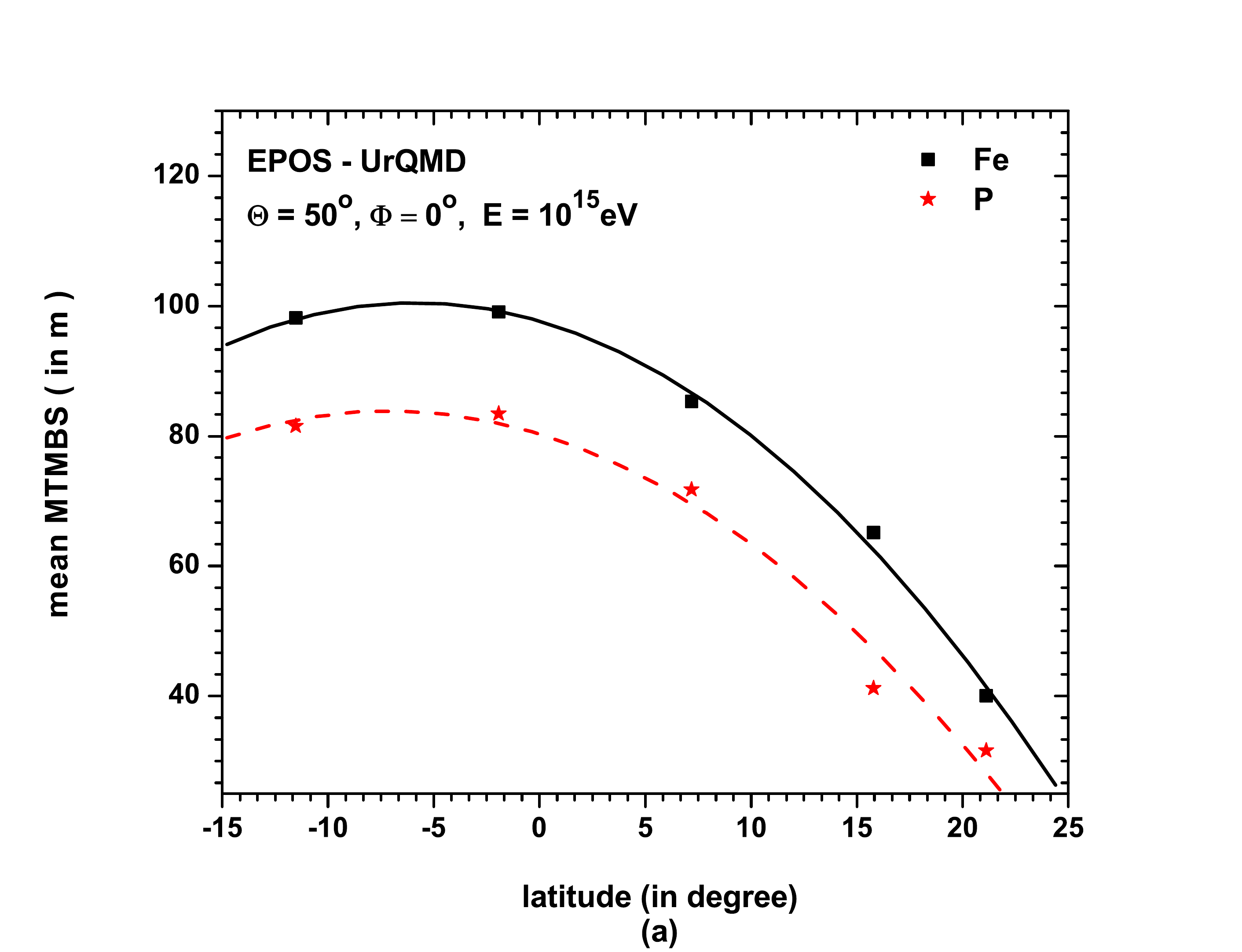}}
\subfigure 
{\includegraphics[scale=0.25]{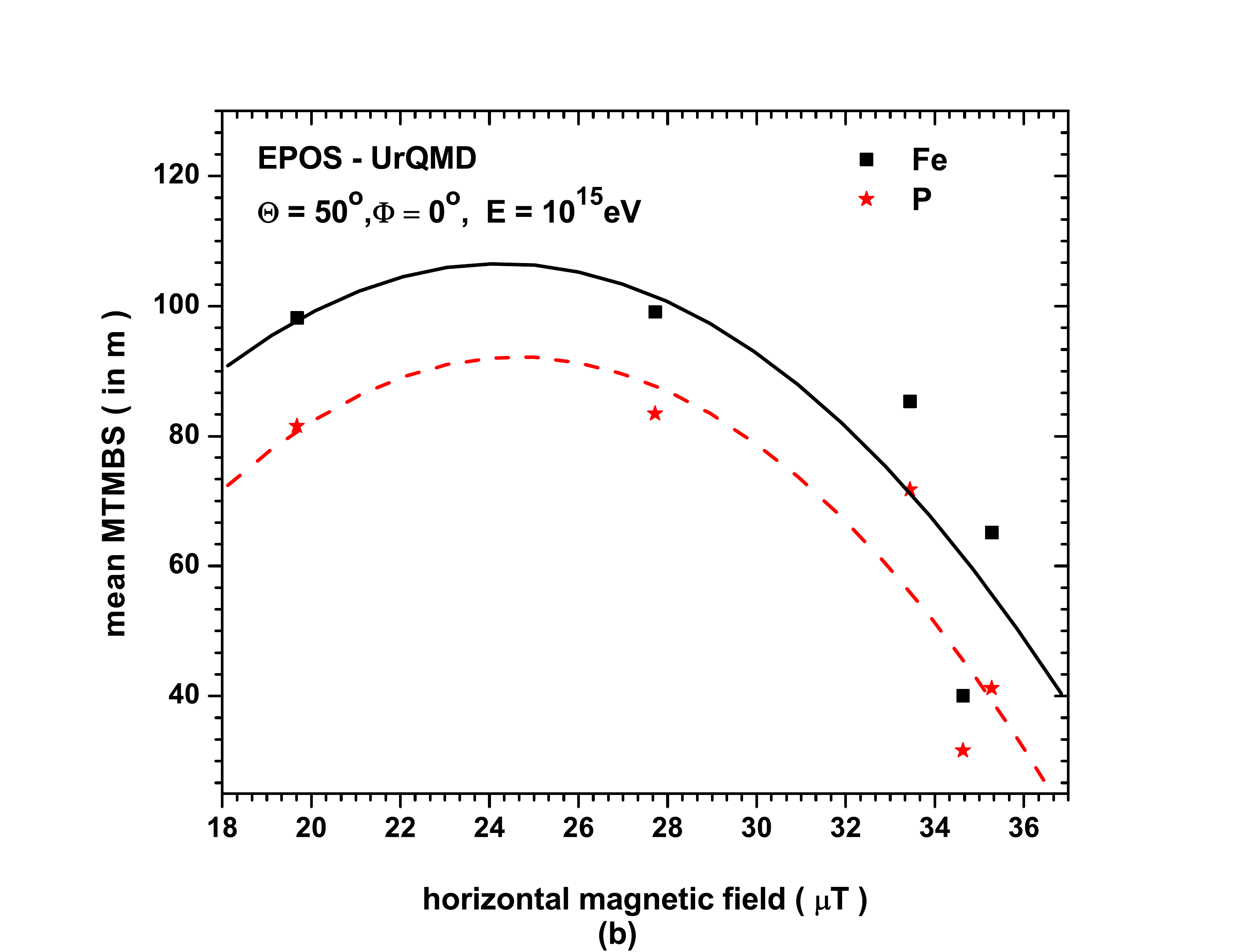}}
\\
\subfigure
{\includegraphics[scale=0.25]{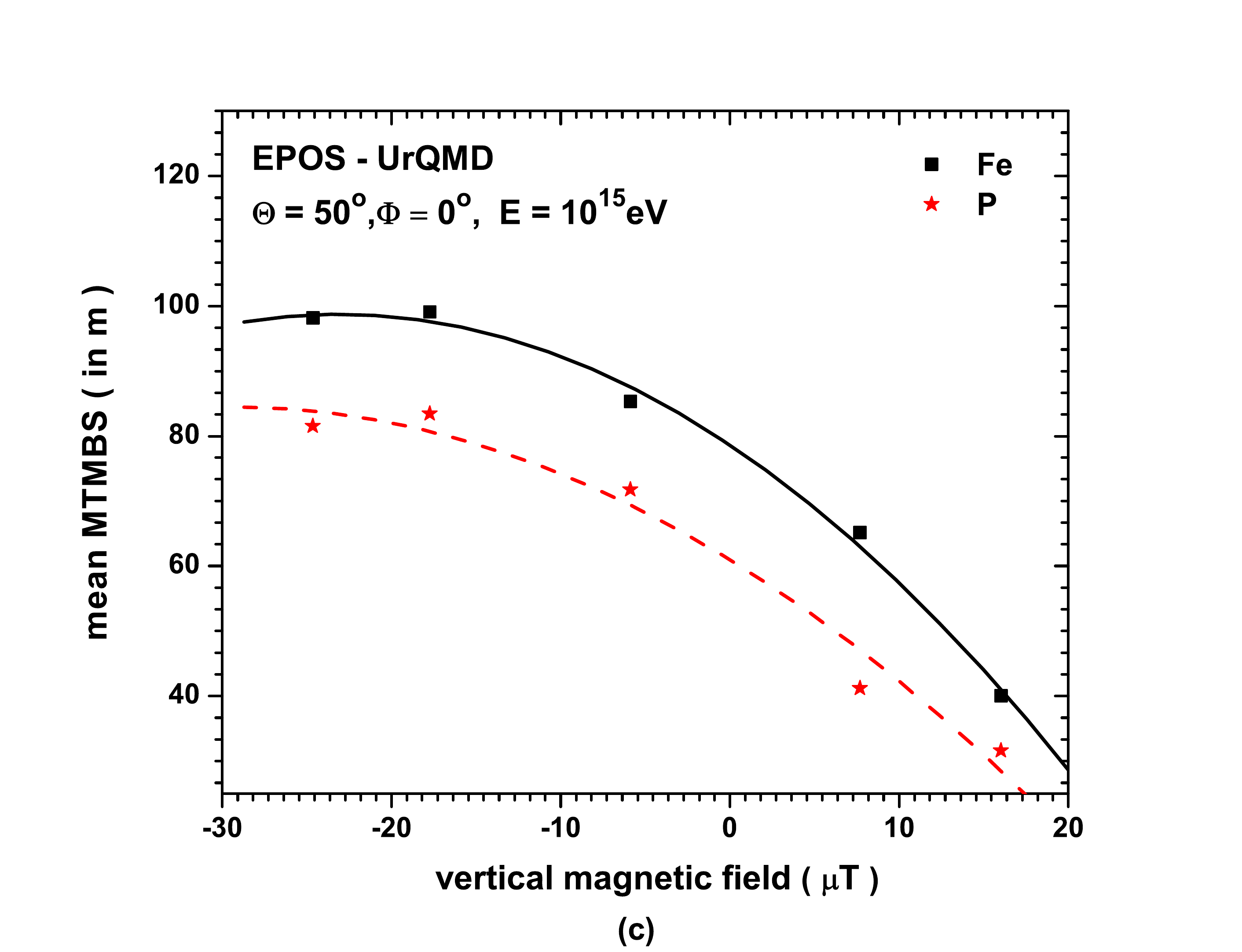}}
\subfigure 
{\includegraphics[scale=0.25]{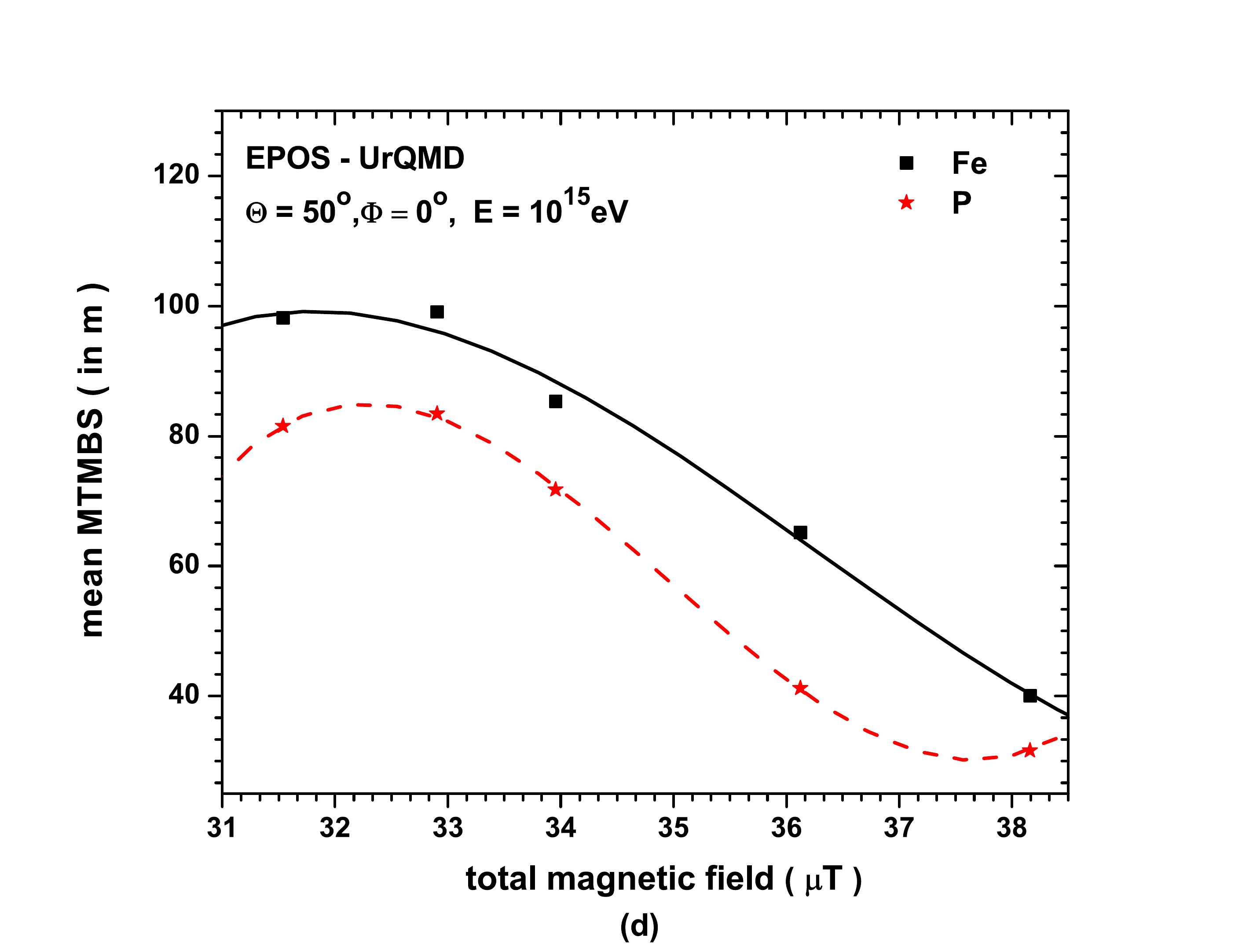}}

\caption{Dependence of the mean MTMBS parameter on latitude and GMF components for locations given in Table II. The solid and dotted curves show these variations according to the best-fit with polynomial functions up to second order in the parameter representing the X-axis.} 
\end{figure*} 
\begin{figure*}[ht]
\subfigure
{\includegraphics[scale=0.25]{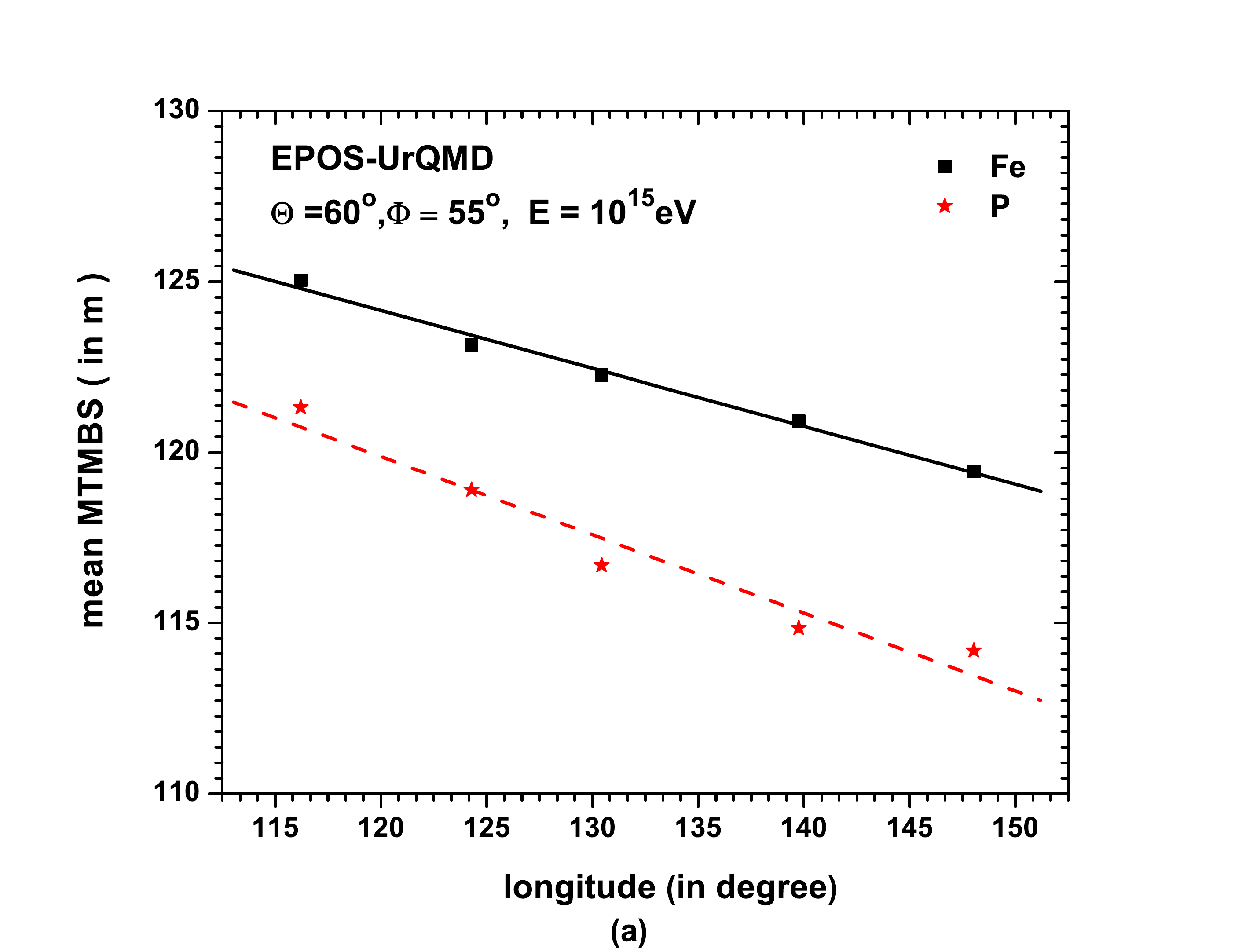}}
\subfigure 
{\includegraphics[scale=0.25]{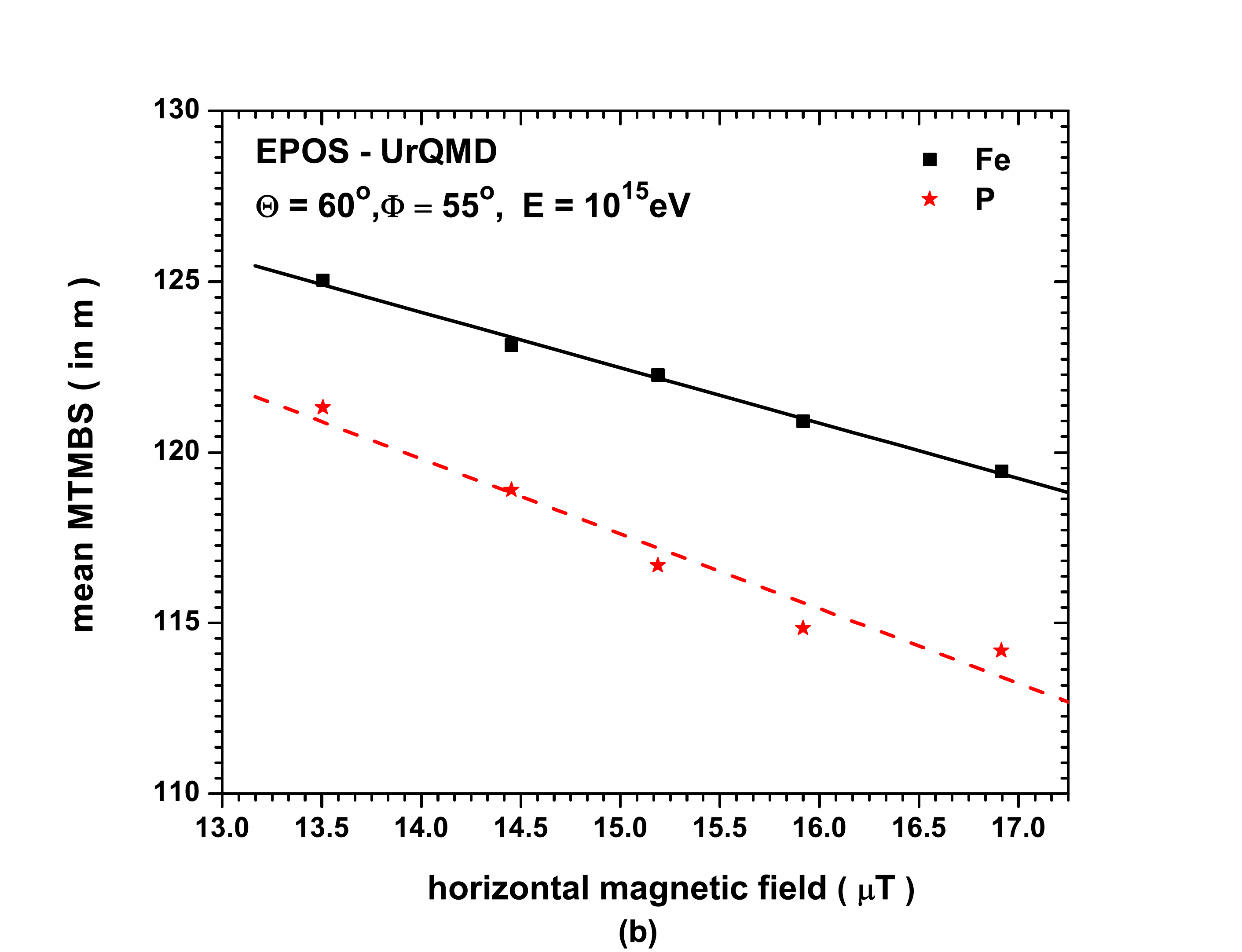}}
\\
\subfigure
{\includegraphics[scale=0.25]{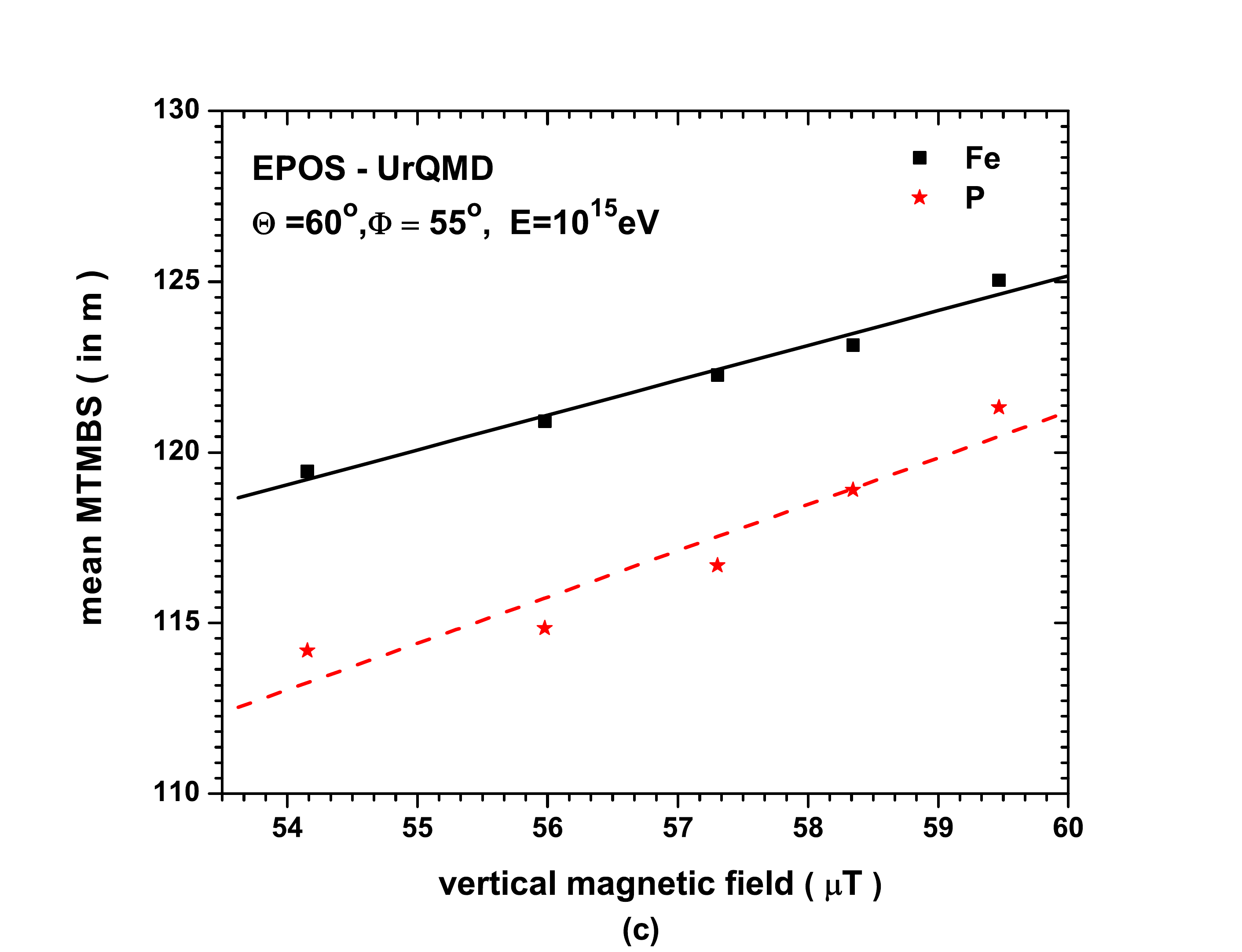}}
\subfigure 
{\includegraphics[scale=0.25]{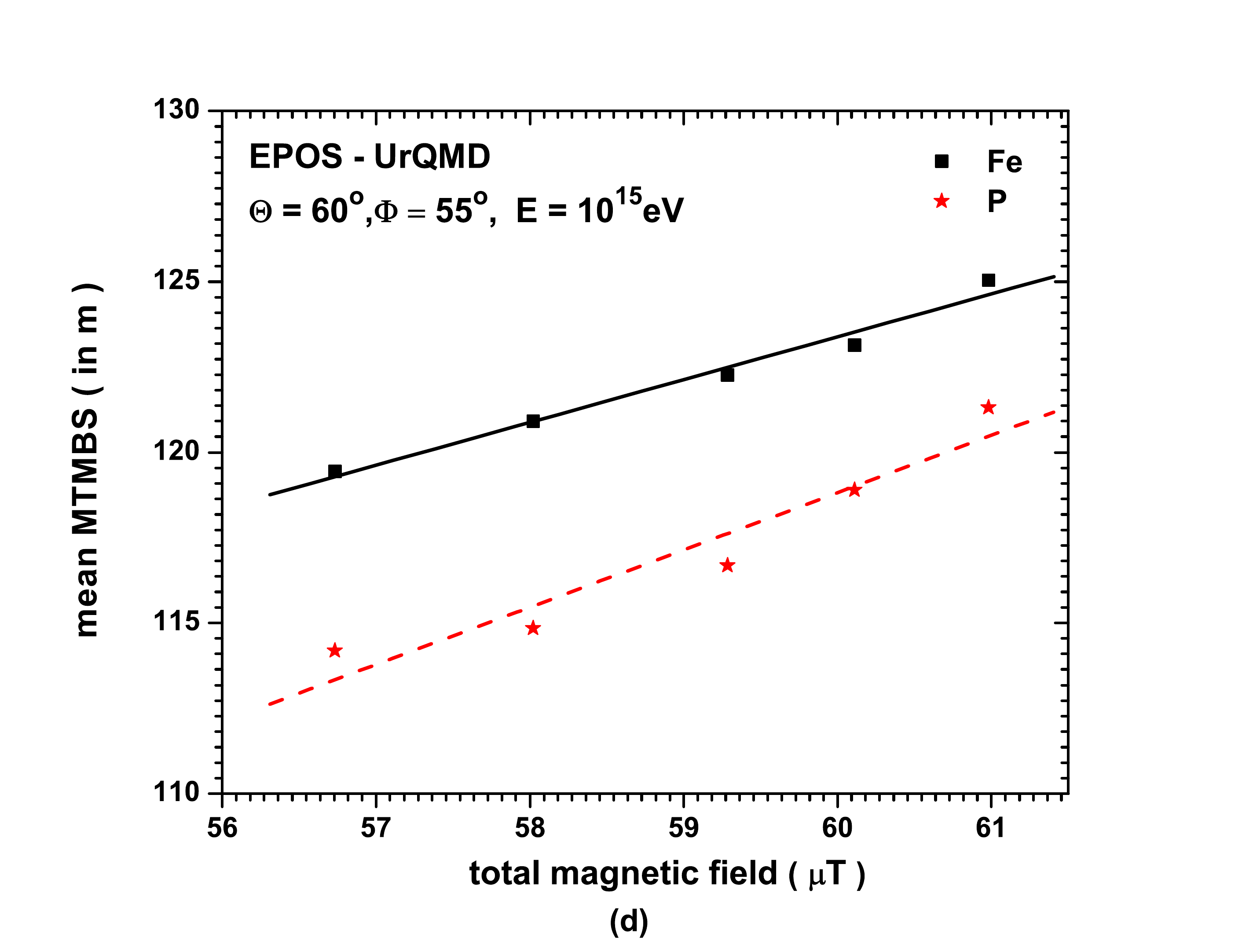}}

\caption{Dependence of the mean MTMBS parameter on longitude and GMF components for locations given in Table I. The solid and dotted lines show these variations according to the best-fit with straight lines.} 
\end{figure*}

We have chosen geographical sites from both the tables I and II for searching some sort of correlations of the GMF components with the MTMBS parameter. We have generated about $10^{3}$ MC showers for each location using necessary information from NOAA. 

As we know that the polar distributions of $\mu^{+}$ and $\mu^{-}$ are affected due to the GMF, and hence any variation in the GMF components should alter the MTMBS parameter. It is also known that the GMF does not vary considerably with time but its intensity is not uniform across the globe with different latitude and/or longitude. In Fig. 8a, 8b and 8c, we have shown the variation of the estimated MTMBS parameter respectively with the total GMF and the GMF components of five locations (Table II). Similar kind of studies have also been carried out for the geographical locations according to the Table I with varying GMF components, and are displayed in Fig. 9a, 9b and 9c. The dependence of the MTMBS parameter particularly on GMF components of different locations indicating some sort of correlations of the MTMBS parameter arising out of the asymmetric $\mu^{+}$ and $\mu^{-}$ distributions with various GMF components.

\begin{table*}
\begin{center}
\begin{tabular}
{|l|l|l|l|l|l|r|} \hline
Zenith & Azimuth & $\rm {B_{H}}$ & $\rm {B_{V}}$ & $\Delta{\rm {B_{V}}}$ & MTMBS & Error\\ 
angle&angle&($\mu\rm T$)&($\mu\rm T$)&($\mu\rm T$)&(m)&(m)\\
\hline
$65^{o}$ & $52.5^{o}$ & 20.52  & 43.57 & 0.00  & 113.023  & 1.537 \\ 
$65^{o}$ & $52.5^{o}$ & 20.52  & 42.87 & 0.70  & 113.684  & 0.577\\
$65^{o}$ & $52.5^{o}$ & 20.52  & 42.57 & 1.00  & 112.915  & 0.532 \\ 
$65^{o}$ & $52.5^{o}$ & 20.52  & 41.57 & 2.00  & 112.259  & 0.616\\
\hline   							
\end{tabular}
\caption {Simulation results at KASCADE level with reduced GMF component (column 4).} 
\end{center}
\end{table*}

\begin{table*}
\begin{center}
\begin{tabular}
{|l|l|l|l|l|l|r|} \hline
Zenith & Azimuth & $\rm {B_{H}}$ & $\rm {B_{V}}$ & $\Delta{\rm {B_{V}}}$ & MTMBS & Error\\ 
angle&angle&($\mu\rm T$)&($\mu\rm T$)&($\mu\rm T$)&(m)&(m)\\
\hline
$65^{o}$&$ 52.5^{o} $&22.22  & 41.99 & 0.00  & 109.743  & 0.573 \\ 
$65^{o}$&$ 52.5^{o} $&22.22  & 41.29 & 0.70  & 107.944  & 0.558\\
$65^{o}$&$ 52.5^{o} $&22.22  & 40.99 & 1.00  & 107.552  & 0.559 \\  
$65^{o}$&$ 52.5^{o} $&22.22  & 39.99 & 2.00  & 107.360  & 0.608\\
\hline   							
\end{tabular}
\caption {Simulation results at an arbitrarily chosen site with an altitude 1608 m (a.s.l.) and the reduced  GMF (column 4).} 
\end{center}
\end{table*}

\subsection{Solar activity and short-term change in GMF}

The GMF around Earth due to some of its internal mechanisms varies very slowly with time. However, during solar activity e.g. coronal mass ejection (CME), a short-lived change in GMF is observed [21-22]. This sudden geomagnetic storm must affect the distribution of energetic muons in an EAS and hence to the MTMBS parameter. Due the solar activity, the interplanetary magnetic field (IMF) surges and directs oppositely to the GMF, and hence the GMF intensity weakens.

To investigate the effect of a sudden weakening in GMF during CME on high-energy EAS muons (actually on MTMBS), we have reduced the vertical component of the GMF ($\rm B_{V}$) in the \emph{CORSIKA} steering file in three situations; (i) 0.70$ \mu\rm T$, (ii) 1.00$ \mu\rm T$ and (iii) 2.00$\mu\rm T$, keeping the horizontal component unchanged. These order of reductions in $\rm {B_{V}}$ are comparable to the CME occurred on 22 June 2015 18:40 UT [23]. The type IV burst during the 21 June CME from the sunspot region arrived Earth on 22 June 2015 18:40 UT and triggered a major G4-level storm. The parameters such as the solar wind speed, magnetic field and its vertical component characterizing the CME are available on OMNIWeb [24]. The CME causes a surge of amount 0.04 $\mu{T}$ to the interplanetary magnetic field (IMF). As a consequence the GMF has undergone a reduction through an amount of $\approx 0.7$ $\mu{T}$. The TMBS/MTMBS parameter estimated employing the azimuthal distribution of muons with $p_{\mu}=100 - 1000$ GeV/c serve as an imprint of the impact of the CME on the PeV CRs.

We have simulated only iron showers at the KASCADE level ($110$ m a.s.l.) for investigating the transient phenomenon with $\rm{E} = 98 - 102$ PeV and reduced $\rm B_{V}$ in the CORSIKA steering file by above mentioned magnitudes. To investigate the altitude dependence of the MTMBS parameter, we have also performed simulations by adopting the above procedure at an arbitrarily chosen site with altitude 1608 m (a.s.l.) and slightly different GMF components compared to the KASCADE site. Our results from these studies are displayed in Table V and Table VI respectively. A significant variation, though very little, is noticed in the MTMBS parameter due to the transient weakening of the GMF caused by the above changes in $\rm {B_{V}}$. We have also noticed that the effect of the solar activity is slightly more at the site having higher altitude relative to the KASCADE.

\begin{figure}
\centering
\includegraphics[width=0.5\textwidth,clip]{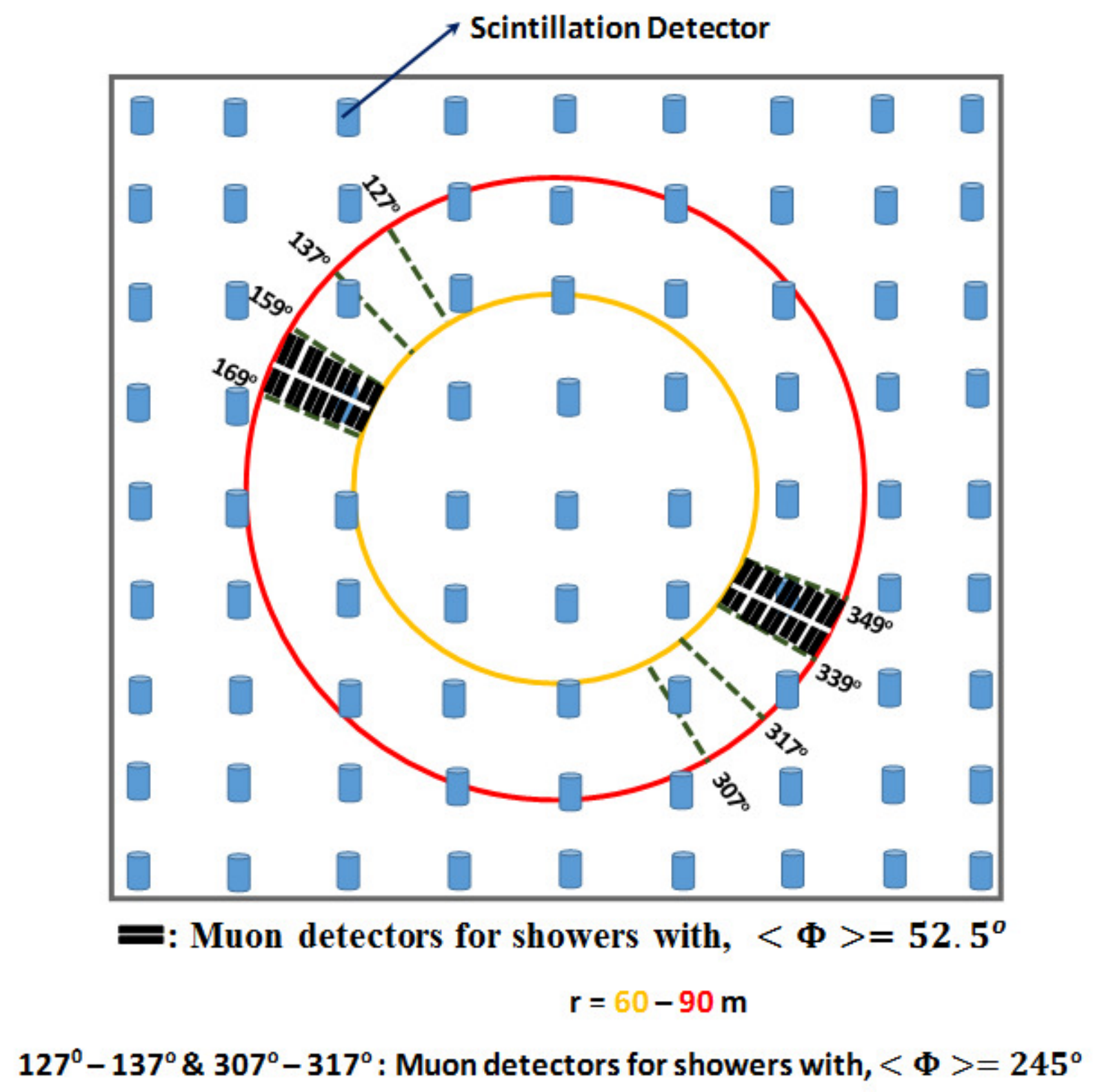} \hfill 
\caption{A possible array layout containing number of scintillation detectors and two muon detecting systems for showers coming from $\langle \Phi\rangle \simeq 52.5^{\rm o}$.}
\end{figure} 

\section{A possible new experiment and summary} 

In this paper we have presented an analysis of the asymmetric muon distributions in highly inclined showers under the effect of Earth's GMF. This analysis has unveiled the expected asymmetry of muons as a function of the observable, TMBS/MTMBS. Furthermore, the employment of the EAS observables such as, $\Theta$, $\Phi$ and the EAS core ($x_{\rm o}$,$y_{\rm o}$) obtained from the lateral distribution of electrons, is essentially important. Nonetheless, the measurement of the sign of the muon charge in an EAS is a foremost task for any possible new experiment.

The measurement of the muon charge sign is the primary concern in any concept of designing a possible experimental set up for implementing the proposed method in the work. The information of the point of impact ($\rm x$, $\rm y$) of each muon in the active volume of a muon detecting system is also necessary. A non-magnetized muon detecting system consisting of shielded scintillators appears robust to detect the charge sign of muons by measuring the life time of the muons, stopped in a stack unit of vertically arranged active (plastic scintillator sheets) and passive (aluminum plates) layers. The possible EAS array should have two muon detecting systems containing several such stack units placed at two selected diametrically opposed regions. The required number of alternatively arranged active and passive layers, and the thickness of the ${\rm e}/{\gamma}$ absorbing topmost layers (concrete/lead) of shielding to detect muons with $\rm{E}_{\mu}^{\rm{Th.}}\geq 100$ GeV can be fixed by the detector simulations. 
   
The mean lifetime of ${\mu}^{-}$ is quite smaller than the mean lifetime of ${\mu}^{+}$ in the present detector material. The method of delayed coincidences is used to distinguish ${\mu}^{+}$ and ${\mu}^{-}$ which are brought to rest in the system. The ${\mu}^{-}$ is captured by the aluminum, and carbon (plastic) atoms to produce a muonic atom that subsequently undergoes successive decay and capture within a short time. However, the ${\mu}^{+}$ particles undergo free decay only and last longer duration than ${\mu}^{-}$. Hence, a significant difference in the mean lifetime of ${\mu}^{-}$ in aluminum compared to that of the free decay of ${\mu}^{+}$, is registered by some simple time electronics with reasonably high rates [5].

The discrete stack units of each muon detecting system give the $\rm x$, $\rm y$ coordinates of a detected muon. The midpoint coordinate of a line connecting the two diagonally placed photomultiplier tubes in the active element is used as the position of detection of a muon. The two muon detecting systems are rotatable to be directed in a particular direction to measure the incoming muons corresponding to different $\Theta$ and $\Phi$. The mid point coordinates of all the diagonals connecting a pair of photomultiplier tubes in an active element are provided from detector simulations for different combinations of $\Theta \geq 55^{\rm o}$ and $\Phi \sim 52.5^{\rm o}$ or $\sim 245^{\rm o}$ in particular. With the help of the simulations, a weight average of all the positive and negative muons from both the muon detecting systems in an experiment can be estimated separately.
\begin{table*}
\begin{center}
\begin{tabular}
{|l|l|l|l|r|} \hline
Zenith & MTMBS & Error & MTMBS & Error\\ 
angle&(m)&(m)&(m)&(m)\\
\hline
$56^{o}$&84.15&$\pm 0.52$&91.43&$\pm 2.08$\\
$57^{o}$&87.18&$\pm 2.41$&93.18&$\pm 0.39$\\
$58^{o}$&88.70&$\pm 1.38$&96.78&$\pm 1.89$\\
$59^{o}$&90.81&$\pm 1.06$&101.50&$\pm 0.53$\\
$60^{o}$&94.17&$\pm 0.39$&101.09&$\pm 0.39$\\
$61^{o}$&98.34&$\pm 1.81$&103.43&$\pm 0.67$\\
$62^{o}$&100.50&$\pm 2.21$&106.18&$\pm 0.81$\\
$63^{o}$&104.73&$\pm 1.43$&107.78&$\pm 1.04$\\
$64^{o}$&106.68&$\pm 0.94$&112.50&$\pm 2.02$\\
$65^{o}$&110.17&$\pm 0.90$&113.02&$\pm 1.54 $\\
\hline   							
\end{tabular}
\caption {Analysis showing zenith angle dependence of the MTMBS parameter for $\Phi =52.5^{o}$ and $\rm {E} = 100$ PeV. Column 2 and 3 stand for proton while column 4 and 5 for iron showers.} 
\end{center}
\end{table*}

\begin{table*}
\begin{center}
\begin{tabular}
{|l|l|l|l|r|} \hline
Azimuth & MTMBS & Error & MTMBS & Error\\ 
angle&(m)&(m)&(m)&(m)\\
\hline
$5^{o}$&66.16&$\pm 0.33$&69.33&$\pm 1.63$\\
$15^{o}$&64.91&$\pm 1.14$&73.43&$\pm 0.61$\\
$25^{o}$&66.08&$\pm 0.92$&75.85&$\pm 2.07$\\
$35^{o}$&73.68&$\pm 1.49$&82.34&$\pm 1.04$\\
$45^{o}$&75.74&$\pm 1.56$&86.52&$\pm 1.16$\\
$55^{o}$&82.74&$\pm 0.42$&91.55&$\pm 0.94$\\
$65^{o}$&86.31&$\pm 0.47$&94.54&$\pm 1.14$\\
$75^{o}$&90.40&$\pm 1.52$&96.34&$\pm 3.34$\\
$85^{o}$&92.81&$\pm 1.17$&101.53&$\pm 0.39$\\
$95^{o}$&94.53&$\pm 1.07$&103.04&$\pm 1.26$\\
$105^{o}$&100.67&$\pm 0.71$&103.10&$\pm 0.81$\\
$115^{o}$&100.67&$\pm 0.71$&107.25&$\pm 1.27$\\
$125^{o}$&100.68&$\pm 1.02$&105.01&$\pm 1.33$\\
$135^{o}$&101.54&$\pm 1.56$&108.17&$\pm 1.44$\\
\hline   							
\end{tabular}
\caption{Analysis showing azimuthal dependence of the MTMBS parameter for $\theta = 55^{o}$ and $\rm {E} = 100$ PeV. Column 2 and 3 for proton while column 4 and 5 for iron showers.} 
\end{center}
\end{table*}

Furthermore, the remaining conventional EAS observables like $\Theta$, $\Phi$, ($\rm{x}_{\rm o}$,$\rm{y}_{\rm o}$), and $\rm N_{{e}^{\pm}}$ or $\rm{E}$ can be obtained from the information of electron densities and their arrival times. For this purpose the experimental setup should have a densely packed array of scintillation detectors in association with the above pair of muon detecting systems. A possible experimental layout containing an EAS array of scintillation detectors and a pair of muon detector units covering two diagonally opposed regions; $159^{\rm o} - 169^{\rm o}$ and $339^{\rm o} - 349^{\rm o}$ is shown in the Fig. 10 for $\langle \Phi\rangle \simeq 52.5^{\rm o}$. The other IQS position in the figure between the pair of dotted lines for each, in opposed positions ; $127^{\rm o} - 137^{\rm o}$ and $307^{\rm o} - 317^{\rm o}$ is displayed for showers with $\langle \Phi\rangle \simeq 245^{\rm o}$.

 If an EAS array contains closely packed scintillation detectors, the core of a shower can be measured very accurately, which is inevitable to the more accurate estimation of $\rm{x}$, $\rm{y}$ coordinates of muons. With reference to the MTMBS parameter, an indication
of the required resolution of the array (e.g. KASCADE) can be known. An estimation of the $\Theta$ resolution can be done by generating about 20 showers each for p and Fe primaries at fixed $\Phi = 52.5^{\rm o}$, $\rm E = 100$ PeV and $\Theta = 55^{\rm o}$. Keeping $\Phi = 52.5^{\rm o}$ and $\rm E = 100$ PeV unchanged, we have simulated showers also for cases with $\Theta = 56^{\rm o}$, $57^{\rm o}$......$65^{\rm o}$. For each $\Theta$ the corresponding MTMBS parameter has been estimated. We have observed that if two p showers maintain a minimum of $\geq 2^{\rm o}$ $\Theta$ difference ($\delta\Theta$) then the MTMBS parameter could clearly discriminate them including errors even at fixed $\Phi$ and $E$. For Fe showers we have got the same resolution for $\Theta$ at these conditions. Under the same conditions, the MTMBS parameter may discriminate between p and Fe showers with $\delta\Theta \leq 1^{\rm o}$. We have applied the same procedure for the estimation of the required $\Phi$ resolution. Here $\Theta$ and $E$ are fixed at values $55^{\rm o}$ and 100 PeV but $\Phi$ varies as $5^{\rm o}$, $10^{\rm o}$,.....$135^{\rm o}$ in each case. The resolution in $\Phi$ for the same type of primaries i.e. either p or Fe is closer to an average value $\simeq 7.5^{\rm o}$ while for p and Fe it comes out as much smaller than $7.5^{\rm o}$. A summary of the angular resolution study of the proposed array using the MTMBS parameter is shown in the Table VII and Table VIII. 

In the Fig. 10, each smaller black top surface refers a lead shielding above a WILLI detector type muon detecting unit which is set at the threshold energy 100 GeV for muons. It was noticed that the distinctive features appearing in Fig. 4 require muons with energies nearly 100 GeV (also noticed in Table IV). This was the reason for considering the high energy muons in the work. Now, generally a magnetic spectrometer associated with tracking detectors could be used to spectroscopy positive and negative muons in order to determine the charge sign from the curvature of the tracks of the incident muons. The present work uses the important observable TMBS which appears due to the influence of Earth's magnetic field on positive and negative muons. Hence, the employment of a magnetic spectrometer under the EAS array could be a possibility. In such a scheme the factors like size, cost and technology towards the spectrometer designing are very crucial. Instead, the shielded scintillators could be a more better option to spectroscopy muon charges. Here, the life times of the muons stopped in a stack of scintillators are used to identify the charge sign of muons. We can use huge blocks of concrete and lead as rooftops of muon detectors or even underground muon detectors buried in mines to make these detectors compatible with a rotating system. Therefore, a future experimental set up consisting of underground rotating muon detecting system and an EAS array on Earth surface seems very challenging towards the implementation of the proposed method.

The proposed method gave evidence to determine the primary CR mass composition at least in the energy region $1 - 100$ PeV. A correlation of certain short-lived weakening of the GMF due to solar activity (\emph{viz.} CME) with the corresponding asymmetries in $\mu^{+}$ and $\mu^{-}$ distributions of a CR induced EAS has been noticed.
 
\section*{Acknowledgment}
RKD acknowledges the financial support from the SERB, Department of Science and Technology (Govt. of India) under the Grant no. EMR/2015/001390. We thank A. Basak for his help in running simulations.

\end{document}